\documentclass[journal]{IEEEtran}
\usepackage{graphicx} %package to manage images
\usepackage{multirow}
\usepackage{subfigure}
\usepackage{color}
\usepackage{hhline}
\usepackage{xcolor}
\usepackage{transparent}
\usepackage{url}
\usepackage{graphics}
\usepackage{graphicx}
\usepackage{subfigure}
\usepackage{textcomp}
\usepackage{booktabs}
\usepackage{siunitx}
\usepackage{multirow}
\usepackage{multicol}
\usepackage{cite}
\usepackage{color, soul}
\usepackage{comment}

\usepackage{algorithm}
\usepackage{algpseudocode}
\usepackage{amsmath}
\usepackage{makecell}

\begin{document}
% ------
\title{Improving Speech Enhancement Performance by Leveraging Contextual Broad Phonetic Class Information}
%
% Single address.
% ---------------

%\author{Authors }% <-this % stops a space
\author{Yen-Ju Lu, Chia-Yu Chang, Cheng Yu, Ching-Feng Liu, Jeih-weih Hung, Shinji Watanabe,~\IEEEmembership{Fellow,~IEEE}, 

Yu Tsao,~\IEEEmembership{Senior Member,~IEEE}

\thanks{Yen-Ju Lu, Chia-Yu Chang, Cheng Yu, and Yu Tsao are with the Research Center for Information Technology Innovation, Academia Sinica, Taipei, Taiwan, corresponding e-mail: (yu.tsao@sinica.edu.tw).}
\thanks{Ching-Feng Liu is with the Chi Mei Hospital.}
\thanks{Jeih-weih Hung is with the Department of Electrical Engineering at National Chi Nan University.}
\thanks{Shinji Watanabe is with School of Computer Science Carnegie Mellon University, USA.}
\thanks{* The first two authors contributed equally to this work }
}

%\thanks{Jeih-weih Hung is with the Department of Electrical Engineering at National Chi Nan University.}
%\thanks{Shinji Watanabe is with School of Computer Science Carnegie Mellon University, USA.}
%\thanks{Yen-Ju Lu, Chia-Yu Chang, Cheng Yu, and Yu Tsao are with the Research Center for Information Technology Innovation, Academia Sinica, Taipei, Taiwan, corresponding e-mail: (yu.tsao@sinica.edu.tw).}

%\markboth{}%

% make the title area
\maketitle
\begin{abstract} 
Previous studies have confirmed that by augmenting acoustic features with the place/manner of articulatory features, the speech enhancement (SE) process can be guided to consider the broad phonetic properties of the input speech when performing enhancement to attain performance improvements. In this paper, we explore the contextual information of articulatory attributes as additional information to further benefit SE. More specifically, we propose to improve the SE performance by leveraging losses from an end-to-end automatic speech recognition (E2E-ASR) model that predicts the sequence of broad phonetic classes (BPCs). We also developed multi-objective training with ASR and perceptual losses to train the SE system based on a BPC-based E2E-ASR. Experimental results from speech denoising, speech dereverberation, and impaired speech enhancement tasks confirmed that contextual BPC information improves SE performance. Moreover, the SE model trained with the BPC-based E2E-ASR outperforms that with the phoneme-based E2E-ASR. The results suggest that objectives with misclassification of phonemes by the ASR system may lead to imperfect feedback, and BPC could be a potentially better choice. Finally, it is noted that combining the most-confusable phonetic targets into the same BPC when calculating the additional objective can effectively improve the SE performance. 

\end{abstract}

\section{Introduction}
\label{sec:intro}
\IEEEPARstart
Speech enhancement (SE) systems improve the intelligibility and quality of speech signals by searching for mapping between distorted speech signals and their clean counterparts. SE has been widely adopted as a front-end processor in various real-world applications such as assistive listening devices \cite{lai2016deep,wang2017deep}, speech coding \cite{accardi1999modular,martin1999new}, speaker recognition \cite{SE_SR}, and automatic speech recognition (ASR) \cite{li2014overview,weninger2015speech, erdogan2015phase,chao2021tenet,kinoshita2020improving}. With recent breakthroughs in deep learning (DL), DL-based SE methods have been extensively investigated and have exhibited outstanding performance \cite{lu2013speech,xu2014regression, wang2014training,tan2019real,kolbaek2016speech,Qi2019SE,xia2014wiener,liu2014experiments,shivakumar2016perception,hu2020dccrn,qi2020tensor,xian2021convolutional,zhang2020deepmmse}. DL-based SE models with nonlinear processing units can learn high-order statistical information from the denoising process. Accordingly, they can considerably outperform conventional SE methods, particularly in extremely low-SNR scenarios and/or non-stationary noise environments. Furthermore, alternative signal processing approaches allow end-to-end DL-wise neural networks to incorporate speech signals with heterogeneous data. For instance, previous studies have confirmed the effectiveness of leveraging face/lip images \cite{8323326} and symbolic sequences for acoustic signals \cite{Liao2019} in SE systems.

To enhance the performance of SE systems, researchers have explored the use of additional losses, including perceptual losses \cite{germain2018speech,yao2019coarse,hsieh2020improving,yang2022improving} and losses from acoustic models (AM) \cite{kataria2021perceptual,bagchi2018spectral,plantinga2021perceptual}. The former group of studies typically computes losses between clean speech and the corresponding enhanced speech in subsequent networks. For instance, in \cite{germain2018speech}, an audio classification network is constructed, and the feature loss function is computed based on the difference between the feature activations of the clean reference signal and the denoised output in an intermediate layer of the classification network. In \cite{yao2019coarse}, the dynamic perceptual loss is calculated as the difference between the classification results of clean speech and enhanced speech from the discriminative network, which is then used to update the SE model. Moreover, \cite{hsieh2020improving,yang2022improving} employ pre-trained models that provide phonetic and acoustic information of audio signals to compute the loss between the acoustic features of clean speech and enhanced speech.

Due to the high correlation between the enhancement and recognition of speech, phoneme information in acoustic models (AM) has been used to improve SE performance. In \cite{kataria2021perceptual}, a variety of perceptual losses were tested using pre-trained AMs for different tasks, including acoustic event detection, automatic speech, speaker, and emotion recognition. Both\cite{bagchi2018spectral} and \cite{plantinga2021perceptual} add a perceptual loss by passing both clean and denoised spectral features to a pre-trained AM and computing the L2 distance of the respective outputs. With the emergence of end-to-end speech recognition (E2E-ASR), joint training of SE and E2E-ASR has been studied for the development of robust speech recognition systems \cite{fan2020gated, liu2019jointly}. The E2E-ASR loss has also been shown to help improve SE metrics in \cite{subramanian2019speech}. 

These works have demonstrated the benefits of deep features and phonetic features for speech enhancement. However, the performances are limited by the accuracy of phoneme identification task. This motivated our prior work using broad phonetic class (BPC) posteriorgrams \cite{lu2020incorporating}. We also demonstrated that speech signals within the same BPC share the same noisy-to-clean transformation. Moreover, the SE model to be learned may combat the noise effect when the original training set has contextual information (desired redundancy) among speech signals. Based on these observations, the contextual information of BPCs, which was not explicitly used in our earlier work \cite{lu2020incorporating}, will benefit SE. Therefore, we proposed an SE architecture incorporating contextual articulatory information acquired by an end-to-end BPC–ASR system, which is expected to further boost SE performance.

The proposed architecture has an SE model and a BPC-based end-to-end ASR (E2E-ASR) module. The bidirectional long short-term memory (BLSTM) encoder and connectionist temporal classification (CTC)/attention hybrid decoder served as the E2E-ASR module, with its phoneme targets replaced by BPCs. Compared with characters and mono-phonemes, selecting BPCs as labels reduced the prediction difficulty. This study derives two losses from the pre-trained AM, which are used for multi-objective training to train the SE system: the ASR classification loss and the perceptual loss. To compute the losses, we selected ESPnet as the toolkit for the E2E-ASR model and connected it to the output of the SE model to establish an end-to-end SE–ASR system.

%Two kind of loss as to the ASR model are used to train the SE model for getting a better enhancement result. 
The deep-feature embeddings in the ASR model were extracted for perceptual loss training, and the distance between the deep features of clean and distorted speech signals was used as the extra loss for learning the SE model. We considered three types of BPCs: the manner of articulation (BPC(M)), place of articulation (BPC(P)), and data-driven BPC (BPC(D)).

The proposed system was first evaluated on a speech denoising task with TIMIT (English) and TMHINT (Mandarin Chinese) datasets with two standard objective evaluation metrics: perceptual evaluation of speech quality (PESQ) and short-time objective intelligibility (STOI), subjective listening tests, and ASR performance. Further experiments were conducted on the dereverberation task using the TMHINT dataset. Finally, the system was evaluated on an impaired speech task, and a set of subjective evaluations was conducted to test its performance.

%This paper is organized as follows: 
The rest of this paper is organized as follows: Section 2 introduces the criteria used to define the BPC and E2E-ASR systems in ESPnet. Section 3 describes the proposed end-to-end BPC SE–ASR system. Section 4 presents the experimental setup, evaluation results, and analyses. Finally, Section 5 concludes this paper.

\section{Backgrounds}
This section introduces the BPC and end-to-end speech recognition model, which serve as the primary components of our proposed architecture.

\label{sec:format}
\subsection{Broad Phonetic Classes}
BPC categorizes all phonemes into several groups according to the articulation of each phoneme. Acoustic characteristics are similar among phonemes of the same group, such as the manner/place of obstruction of airflow that passes through the mouth. This study adopted three BPC grouping methods: knowledge-based and data-driven. 

\subsubsection{Knowledge-based BPC}
This study used two knowledge-based BPCs: place and manner of articulation. The place of articulation indicates where the air stream is blocked in the vocal tract when a sound is uttered. 
The manner of articulation indicates how the air stream is blocked. Similar characteristics appear in phonemes with the same manner and place of articulation, and the type of articulation manner can be discerned by observing the shape of its waveform \cite{ladefoged2014course}. 
Phonemes with the same manner/place of articulation have similar spectral characteristics and may generate significant confusion when performing ASR \cite{scanlon2007using}. Nevertheless, this problem can be alleviated using contextual articulatory information \cite{SINISCALCHI2014326,articles11265,SINISCALCHI2013148,shahrebabakitransfer}.
%, where phonemes with the same air-blocked position are considered as a group
%lin2018improving

Different native languages have different articulatory characteristics, and divergence of the manner/place of articulation enables them to produce distinct sounds and prosody. This study used the International Pronunciation Alphabets (IPAs) to represent the target sentences in any language and characterize each IPA label into BPCs. Eighty-seven phones in IPAs were clustered into nine articulatory manner classes: vowel, plosive, nasal, trill, trap or flap, fricative, lateral fricative, approximant, and lateral approximant. The vowel class includes diphthongs and semi-vowels, as suggested by \cite{scanlon2007using}. Each language uses only a portion of the IPAs to represent all its phonemes. For instance, in the TIMIT dataset, 60 IPAs were used to represent all phonemes in English, and they were clustered into five groups based on articulation manner: vowels, stops, fricatives, nasals, and silence.

For the place of articulation, we used 10 clusters in Mandarin by classifying 87 IPAs: vowels, bilabial, labiodental, dental alveolar posteroalveolar, retroflex, palatal, velar, ular, pharyngeal, and glottal. Comparatively, only nine clusters have been used in English \cite{ladefoged2014course}. In both manner and place articulations, vowels constituted a distinct group from the others. This is because these two classification criteria mainly focus on consonants, whereas uttering vowels do not block the air stream as much as pronouncing consonants.

\subsubsection{Data-driven BPC}
The similarity between phonemes can also be evaluated in a data-driven manner, derived from the recognition result of a pre-trained AM. In a previous study \cite{lopes2012broad}, a confusion matrix $\textbf{M}$ for phonemes was created, where its entry $\textbf{M}_{ij}$ was defined by the number of events for phoneme $i$ to be misidentified as phoneme $j$. This matrix was assumed to reflect the similarities between each pair of phonemes. When clustering phonemes through the similarity metric, a merging process was performed until the cluster number reached nine, as recommended in \cite{lopes2012broad}.

% Each entry $\textbf{S}_{ij}$, represent the distance between a pair phonemes $i$ and $j$, which is computed by (\ref{Houtgast}):

% \begin{equation}
%   \textbf{S}_{ij} = \textbf{S}_{ji} = \sum_{k\neq i, j} \mathrm{min}(\textbf{M}_{ik},\textbf{M}_{jk})
%   \label{Houtgast}
% \end{equation}

% where $k$ is the phone index and $k \neq i, j$. 

%\label{sssec:subsubhead}

\subsection{CTC/attention E2E-ASR}
\label{sssec:ctc_att}
The applied recognition model adopted two major E2E-ASR implementations: CTC and attention. It provided a single neural network architecture to perform speech recognition in an E2E manner \cite{watanabe2017hybrid}. The attention-based method used an attention mechanism to align acoustic frames and recognize symbols. CTC uses Markov assumptions to solve sequential problems efficiently using dynamic programming. Multi-task learning based on CTC and attention allows E2E-ASR to resolve the misalignment issues encountered in ordinary attention-based E2E-ASR.

Compared to conventional ASR models that require various modules, such as AMs, language models, and lexicons, the E2E network eliminates the need for linguistic resources. It enables an optimization of front-end processors that precede the ASR component in an end-to-end manner. Furthermore, the complexity of the E2E-ASR building process is notably reduced, as it does not require GMM/HMM construction, DNN pre-training, lattice generation, and complex searches during decoding, compared to the conventional ASR. Simplifying a unified deep learning framework in E2E-ASR enables researchers to develop or use an ASR system for new applications without extra efforts, such as collecting data on new languages and preparing their linguistic resources. 

In the E2E-ASR model, a shared BLSTM encoder transforms the input sequence into high-level features and undergoes multi-objective learning. When training the E2E-ASR model, the objective functions for the attention and CTC frameworks were applied to improve the robustness and achieve fast convergence. The CTC objective function is an auxiliary task to train the encoder of the attention model. Compared with the single attention model, combining the forward-backward algorithm in CTC accelerates the process of finding the desired alignment in a monotonic manner and mitigates the prediction from a letter-wise attention objective to a sequence-level CTC objective. Through joint decoding, attention- and CTC-based scores were combined in a one-pass beam search algorithm to obtain the ASR results and further eliminate irregular alignments.

\section{PROPOSED MODEL}
\label{sec:pagestyle}
The proposed model connected and trained the CTC/attention E2E-ASR for ASR and a transformer for SE in an E2E manner to improve SE performance with BPC recognition.

\subsection{SE with AM and E2E-ASR multi-objective training}
\begin{figure}[tp!]

\subfigure[The training approach with DNN-HMM hybrid AM model.]{%
  \includegraphics[clip,width=\columnwidth]{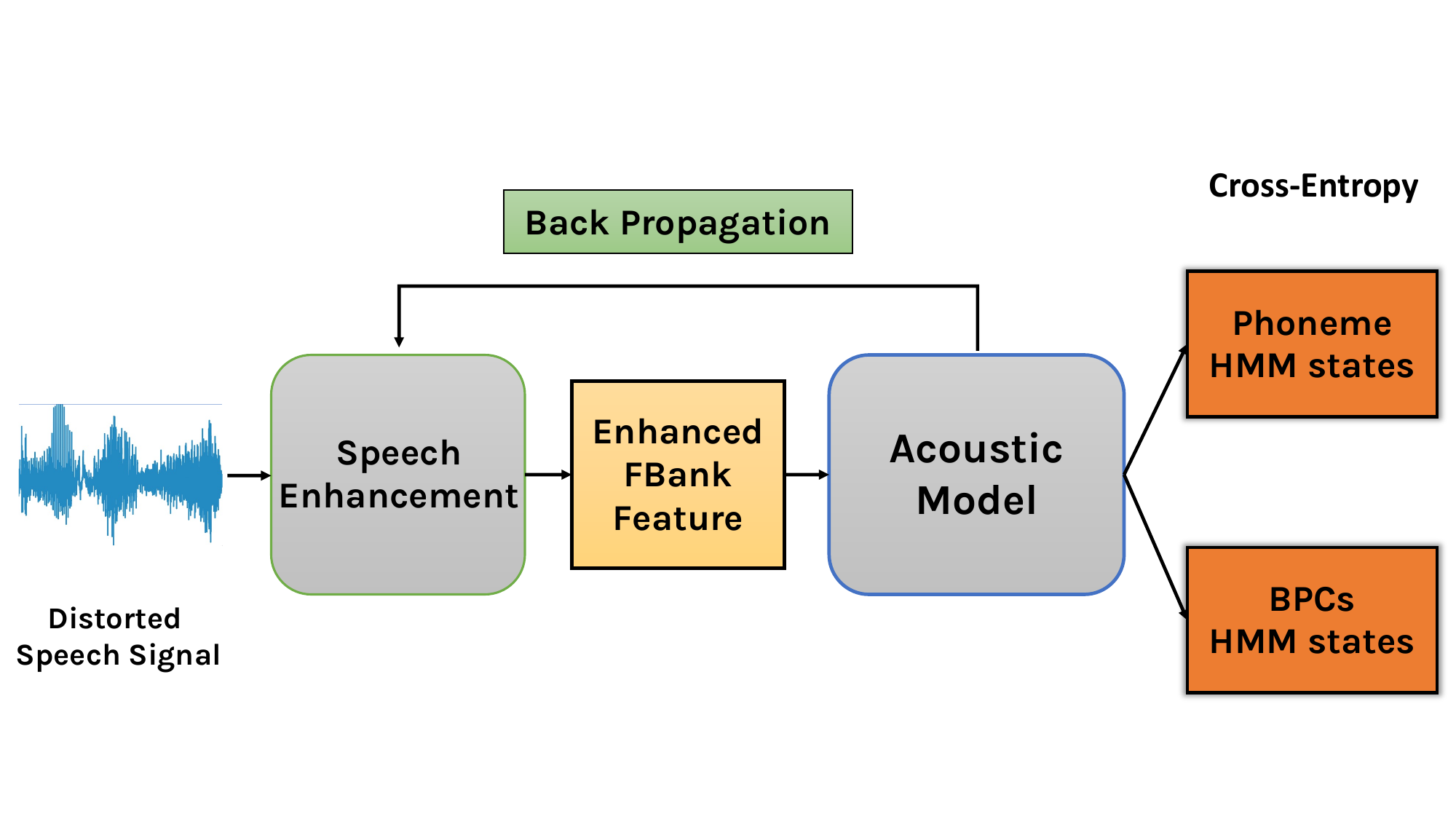}%
  \label{fig:AM}
}

\subfigure[The training approaches with E2E-ASR model.]{%
  \includegraphics[clip,width=\columnwidth]{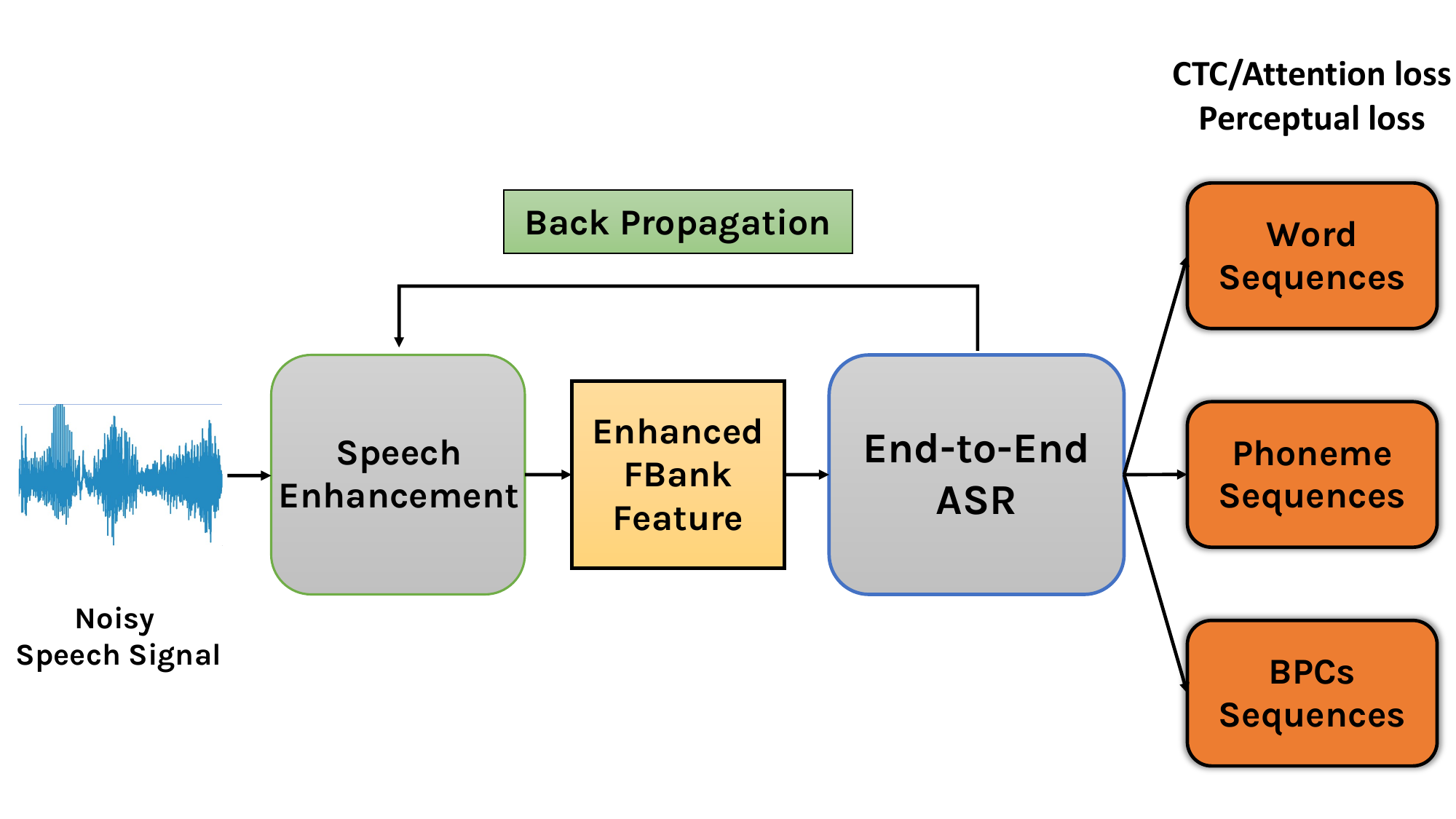}%
  \label{fig:ASR}
}

\caption{The training approach with DNN-HMM hybrid AM model and E2E-ASR model. The training targets of DNN-HMM hybrid AM are phonemes or BPCs, and the training targets of E2E-ASR model are words, phonemes, or BPCs.}
\label{fig:recognition}
\vspace{-4mm}
\end{figure}

We hypothesize that the SE model can be further promoted through learning to generate enhanced speech signals with a more precise transformation guided by BPC information. To validate this hypothesis, we set up two SE systems that are estimated by multi-objective training with two different losses. Figure \ref{fig:AM} connects the SE model with a DNN-based AM from the conventional hybrid DNN-HMM ASR system, and both BPCs and phonemes can be HMM states in the AM. Here, the SE model was independently updated in each time frame because the AM only predicts one HMM state at one forward step without considering the long-term context. By contrast, the newly proposed architecture shown in Figure \ref{fig:ASR} connects the SE model with an E2E-ASR model, predicting consecutive BPC labels instead of phoneme/word sequences. Since the E2E model predicts all the BPC labels at once, the SE model can learn information for a longer period and generate results with better transformation of articulatory information. We also conducted SE-E2E-ASR multi-objective training for the sequences of phonemes and words instead of BPCs for comparison, as shown in Figure \ref{fig:ASR}.

The training of conventional DNN-HMM hybrid acoustic models typically has two steps. First, the GMM-HMM acoustic models for different tri-phones were trained using the expectation-maximization (EM) algorithm, which infers the state emission probabilities in the HMM. We then used DNN instead of GMM and trained it to model the HMM states more precisely. We conducted our experiments based on the TIMIT recipe in Kaldi \cite{povey2011kaldi} and built a four-layered DNN using PyTorch-Kaldi \cite{pytorch-kaldi} with dimensions $[1024, 1024, 1024, \mbox{states}]$, where ``states'' denotes the number of states in the HMM. We connected the DNN-based AM after the SE model and used cross-entropy loss from the predicted results of the HMM states to update the SE model.

\subsection{SE-E2E-ASR model architecture}
\label{sssec:e2e}
\begin{figure}
 \centering
 \includegraphics[width=\linewidth]{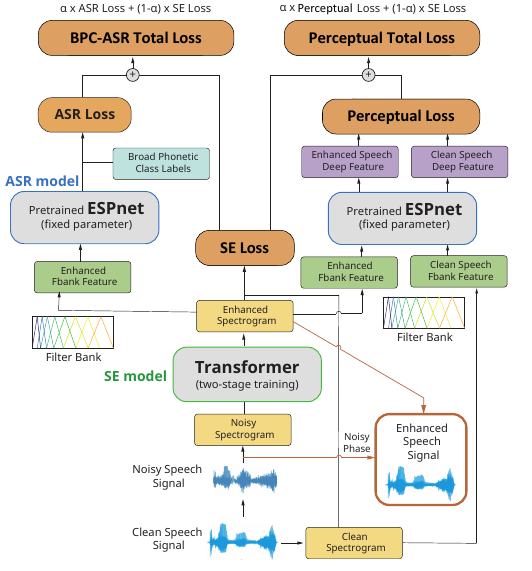}
 \caption{The architecture of the proposed combination of SE model and E2E BPC-ASR Network} 
 \label{fig:E2E}
\end{figure}

Figure \ref{fig:E2E} shows the proposed SE-E2E-ASR architecture, which has a feature extractor, SE model, and BPC-based pre-trained ASR model, which uses the overall loss of both models for back-propagation in training.
The SE model is arranged as a transformer, which is an attention-based deep neural network originally proposed for machine translation \cite{vaswani2017attention} and later applied in numerous other natural language processing tasks. The transformer has shown considerable improvements over common recurrent neural networks (RNNs) and has been further exploited in SE tasks. The E2E-ASR model in ESPnet was initially set to recognize the acoustic waveform into character-level sequences, while its output labels were modified into the desired BPC labels in our ASR model.

\subsubsection{Transformer-based SE model}
Transformers have been investigated extensively in SE studies \cite{hao2019attention, yang2020characterizing, 9053214}. Following a sequence-to-sequence learning structure, the transformer comprises encoder and decoder networks. This method preserved only the encoder part of the transformer for the SE process because the input distorted signal and output enhanced signal share the same length. The transformer has four convolutional layers for encoding the spectrum of the input signals and eight attention blocks. Each attention block comprises multi-headed self-attention and two fully connected layers as the feed-forward network. Residual connections and layer normalization were performed in each layer \cite{ba2016layer}.

\subsubsection{Transformer-based ISE network}
In addition to speech denoising, we also enhanced impaired speech as an alternative SE task. We adopted a voice conversion (VC) model \cite{Huang2020VoiceTN} based on a transformer as the fundamental architecture of the ISE model. The ISE produces recognizable speech from impaired speech. This task applied a sequence-to-sequence (seqASRseq) model based on transformer architecture with text-to-speech (TTS) pre-training. By transferring knowledge from learned TTS models trained by the large TTS dataset, we could satisfy the need for large-scale corpora to train the transformer. This ISE model was primarily based on the transformer-TTS model, consisting of encoder and decoder stacks. The encoder layer has a multi-head self-attention sub-layer followed by a fully connected feed-forward network, and the decoder layer contains another sub-layer, which performs multi-head attention over the output of the encoder stack. Each layer comprises a residual connection and layer normalization. This ISE model takes the source log-Mel spectrogram as input and outputs a converted log-Mel spectrogram. Before training with the ISE objective, the decoder in the transformer was first pre-trained using the TTS-objective tasks. Subsequently, the decoder was fixed to preserve its ability to robustly capture speech features, such as articulation and prosody, and the encoder was trained with TTS speech input to learn the effective hidden representation. Finally, the ISE model was trained by initializing the model using the pre-trained parameters of the encoder and decoder. The models initialized with the TTS-pre-trained model parameters generated effectively hidden representations for high-fidelity and highly intelligible converted speech.

\subsubsection{Mel-filters processing}
\label{sssec:mfp}
To make the whole SE-E2E-ASR system differentiable, we replaced the Kaldi feature extractor in the original ESPnet with a filter-bank extractor, creating speech features of the enhanced waveforms from the SE module. The original ESPnet uses Kaldi feature extraction, for which most recipes use 80-dimensional logarithmic Mel-spectra with the pitch feature (83 dimensions in total). By contrast, the filter-bank extractor applies 26 triangular Mel-scaled filters to the power spectrum of an input signal to extract the filter-bank feature. Compared with the original Kaldi feature extractor, the filter-bank extractor connects the SE module with the ASR model and ensures that the whole SE-E2E-ASR system is differentiable. After feature extraction, we prepared the data for the SE-E2E-ASR system with all the information included in the Kaldi data directory (transcriptions, speaker and language IDs, and input and output lengths) and pre-trained the E2E-ASR model using the clean filter-bank features as input.
The ISE task used the Kaldi feature extractor and preserved the processes in the ESPnet toolkit, where 80-dimensional logarithmic Mel-spectra features (with 1024 DFT points and 144-point frameshift) are used. The Kaldi feature was similar across all inputs and outputs of the ISE model.

% To extract features, we replace the Kaldi feature extractor in ESPnet with filter bank extractor and prepare the data by converting all the information including in the Kaldi data directory (transcriptions, speaker and language IDs, and input and output lengths) to one JSON file. The original ESPnet uses the Kaldi feature extraction, and most of recipes use the 80-dimensional log Mel feature with the pitch feature (totally 83 dimensions). Nevertheless, when training the proposed E2E model, the output enhanced waveforms of the SE model should be extracted into features by differentiable method to keep the whole process back-propagatable. In our method, the filter bank extractor applies triangular filters, typically 40 filters, on a Mel-scale to the power spectrum to extract frequency bands. In comparison with the original Kaldi feature extractor, the filter bank extractor can connect the SE model with the ASR model and ensure the E2E training process. The ESPnet is pre-trained using the clean filter bank features as input before the overall E2E training.

\subsubsection{E2E BPC-ASR Network}
To obtain BPC context information, we changed the original output word labels of E2E-ASR to the desired BPC labels. Accordingly, the BPC–ASR model predicted the BPC label sequence that corresponds to each utterance. We pre-trained the BPC–ASR model using the filter-bank features of clean utterances as the input for the overall E2E model training. In order to avoid the SE model from producing distorted signals while being used as new features for training the ASR model during joint training, the parameters of the pre-trained BPC-ASR model were kept fixed or "frozen" during the training of the connected SE model.

\subsection{Multi-objective training methods}
\label{sssec:jtm}

\subsubsection{SE-ASR multi-objective training}
\label{sssec:seasrjt}
We proposed two scenarios of multi-objective training as follows: 
Optimization of the E2E-ASR objective function was treated as an auxiliary task to train the transformer-based SE model. We combined the losses of both the SE and E2E-ASR models with a tuning parameter $\alpha$ to form the total loss for multi-objective training, as shown in Eq. (\ref{eqASR}).

\begin{equation}
  \textbf{loss}_\text{total} = (1-\alpha) \times \textbf{loss}_\text{SE} +\alpha \times \textbf{loss}_\text{ASR}, 
  \label{eqASR}
\end{equation}

The L1 loss function was used to calculate the SE loss and the combined CTC and attention losses from the E2E-ASR model was used as the ASR loss. Parameter $\alpha$ was tuned to make the two losses (SE and ASR losses) contribute almost equally to the total loss in Eq. (\ref{eqASR}) and was set in the range $[0.001, 0.002]$ because the ASR loss is usually exceedingly larger than the SE loss. Furthermore, since the ASR model was pre-trained, the parameter $\alpha$ was set to $0$ in the first stage of the multi-objective training to learn the SE model alone without considering the ASR loss. This arrangement yielded better SE performance in our preliminary experiments.

\subsubsection{SE optimization integrating perceptual loss}
\label{sssec:dfl}
In addition to the recognition error used as the ASR loss, we exploited the ASR model to create another objective function that considered the perceptual loss to train the SE model. The respective model architectures are shown on the right side of Figure \ref{fig:E2E}. The E2E-ASR model extracted the features in the last layer (320) of the encoder from both clean and distorted utterances, known as deep features. Subsequently, the L1 loss between the clean and enhanced deep features was set as another form of loss ($\textbf{loss}_\text{PL}$), which was combined with the spectrum-wise SE loss ($\textbf{loss}_\text{SE}$) for multi-objective training, as shown in Eq. (\ref{eqPL}). The perceptual loss evaluated the distance between clean and enhanced signals at the layer level inside the ASR model; thus, it is considered as another SE loss for the ASR. During training with the perceptual loss, only the pre-trained ASR model is required without the corresponding transcriptions for the enhancement training data.

\begin{equation}
  \textbf{loss}_\text{total} = (1-\alpha) \times \textbf{loss}_\text{SE} +\alpha \times \textbf{loss}_\text{PL}, 
  \label{eqPL}
\end{equation}  
  where $\alpha$ is a tuning parameter.
  
%Instead of using the recognition loss as the ASR loss in the multi-objective training} stage, we proposed another objective function in ASR model to focus on the speech enhancement performance. Figure \ref{fig:E2E} demonstrates the model architecture of this method. In ESPnet, we extract the features at the last layer (with size 320)of the encoder of both clean and noisy speech and compute loss between them to set enhancement as objective. This enhancement-oriented ASR loss is also added to the SE loss as equation (\ref{eqASR}) in multi-objective training} stage.

\subsubsection{SE optimization integrating three losses}
\label{sssec:coml}
Both the ASR loss ($\textbf{loss}_\text{ASR}$) and the perceptual loss ($\textbf{loss}_\text{PL}$) can be used for training simultaneously. $\textbf{loss}_\text{ASR}$ guides the SE model to generate enhanced speech with better prediction results in the recognition model, while $\textbf{loss}_\text{PL}$ can make the model generate enhanced speech with prediction results that are closer to the clean speech by leveraging a pre-trained ASR. The combined loss of all three losses is:

\begin{equation}
  \textbf{loss}_\text{total} = (1-\alpha_1-\alpha_2) \times \textbf{loss}_\text{SE} +\alpha_1 \times \textbf{loss}_\text{ASR} + \alpha_2 \times \textbf{loss}_\text{PL}, 
  \label{eqCom}
\end{equation}

  where $\alpha_1$ and $\alpha_2$ are the weights for $\textbf{loss}_\text{ASR}$ and $\textbf{loss}_\text{PL}$.

\section{EXPERIMENT}
\label{sec:typestyle}
Two datasets were used to evaluate the proposed architecture: the TIMIT corpus \cite{garofolo1993timit} and the Taiwan Mandarin version of hearing in noise test (TMHINT) sentences \cite{huang2005development}. The following section introduces the experimental setups and presents the evaluation results and respective analyses and discussion.

\subsection{Experiments on the TIMIT dataset}
\label{sssec:timit_exp}
% \label{sssec:subsubhead}

For the experiments conducted on the TIMIT database with multiple noise sources, 10,000 noisy-clean paired training utterances were used, comprising 3,696 utterances in the training set with an average duration of 4 seconds and their noisy counterparts containing 102 noise types from \cite{hu2004100} at six different SNR levels (20, 15, 10, 5, 0, and -5 dB). The core test set of TIMIT (including 192 utterances) was mixed with five unseen noise types at four SNR levels (5, 0, -5, and -10 dB) to build the test set in our experiments. The training and test sets did not share common speakers.

The speech waveforms were recorded at a 16 kHz sampling rate and converted into 257-dimensional spectrograms using the short-time Fourier transform with a Hamming window size of 32 ms and a hop size of 16 ms. The $\mathrm{log1p}$ function ($\mathrm{log1p(x)}=\log(1+x)$) was adopted on the magnitude spectrogram to ensure non-negative outputs \cite{fu2020boosting}, and normalization was performed on the waveform. The test stage combined the enhanced magnitude spectral features and original phases from the noisy signals to synthesize the enhanced signals. 

As previously mentioned, two-stage training was applied to train the SE model. The SE model was first trained for 70 epochs without considering the ASR results (by setting $\alpha=0$ in Eqs. (\ref{eqASR})) and (\ref{eqPL})), and was further updated with the combined objective loss, as in Eq. (\ref{eqASR})) and \ref{eqPL} (by setting $\alpha=0.001$ for the ASR loss and $\alpha=0.05$ for the perceptual loss) for the next 80 epochs. For the experiment that combines all three losses, we have reduced the weightage of $\alpha$ to half for both the ASR loss and perceptual loss. Specifically, we set $\alpha_1=0.0005$ and $\alpha_2=0.025$. This was done to balance the contribution of each loss and ensure that the SE model is trained to produce enhanced speech that is not only better for ASR tasks but also closer to clean speech. The ASR model was pre-trained with the clean dataset, and its parameters were fixed (frozen) when learning the SE model. To prevent overfitting, we performed early stopping based on validation performance. We used the Adam optimizer \cite{kingma2014adam} with a fixed learning rate of $5\times10^{-5}$ for all of our experiments. The SE model and ASR model had sizes of 33.8MB and 30.4MB, respectively. Our experiments were conducted on a GeForce RTX 2080 Ti, with an average training time of 0.88 milliseconds per frame. Please note that the ASR model was only used in the training phase but not in the testing phase.

Based on the model architecture illustrated in Fig. \ref{fig:E2E}, we implemented three systems using three types of BPCs, namely BPC(M), BPC(P), and BPC(D), as the acoustic units in the ASR model.  In the following, we refer SE by leveraging AM only and E2E-ASR as SE-AM and SE-E2E-ASR, respectively.

% \begin{figure}[h]
%  \centering
%  \includegraphics[width=0.98\linewidth]{figures/error_rate.pdf}
%  \caption{The error rates at different SNR levels obtained by the E2E-ASR for monophone and BPCs for noisy and enhanced speech.} 
%  \label{fig:Recognition Result}
% \end{figure}

% \subsection{TIMIT Results and Discussions}
% \label{sssec:subsubhead}

% \subsubsection{Phonetic/BPC recognition error rates}
% To corroborate the assumption that the E2E-ASR model can guide the SE process, we first analyzed the error rates of the E2E-ASR system in noisy and enhanced conditions, as shown in Figure\ref{fig:Recognition Result}. First, we observed that the error rate for the monophone system increased significantly as the SNR decreased, reaching 84$\%$ at 0 dB SNR and exceeding 100$\%$ at $-10$ dB SNR. Next, compared with the results for the monophones, the error rates for the BPCs increased less significantly with the noise level (e.g., the error rates were from 37$\%$ to 52$\%$ for BPC(M)). Subsequently, we discovered that the BPC(D), which employed the confusion matrix from the AMs, had the lowest error rate; this was likely due to its clustering method. Finally, all of the recognition results improved through SE, indicating that the SE process guided by the contextual phonetic information can benefit the recognition accuracy of E2E-ASR. 

\subsubsection{Results for the Multi-objective Training Models}
Table \ref{tab:am-asr table} presents the PESQ and STOI scores of some baseline models and the proposed SE scenario. In the following experiments, unless otherwise specified, we train SE systems using a combined E2E-ASR and L1 loss (Eq. (\ref{eqASR})) In addition to the unprocessed baseline (denoted as Noisy), we considered the transformer-based SE without ASR loss and perceptual loss as the first advanced baseline for comparison (denoted as SE baseline). Subsequently, the SE-AM applied the multi-objective training of the SE model and the DNN-HMM-based AM model with phoneme, BPC(M), and  BPC(D) as the acoustic unit individually. For the proposed SE-E2E-ASR system, three kinds of acoustic units, namely word, phoneme, and BPC(D), were individually used for the E2E-ASR module to evaluate the corresponding SE results.

% we we mainly use the articulatory contextual information to improve the SE model. We consider the Transformer results without involving multi-objective training} as our first baseline in this study. The SE-AM baselines utilize the multi-objective training} of the SE model and a DNN-based AM model. For our proposed method, the SE-E2E-ASR system, three kinds of acoustic units are individually used: word, phone and BPCM, to evaluate the corresponding SE results.

\begin{table}[t]
\centering
\caption{Averaged PESQ and STOI scores for SE-AM systems with BPC(M) and phoneme, and SE-E2E–ASR systems with BPC(M), phoneme, and word. The scores for the unprocessed and SE baseline are listed for comparison. }
\label{tab:am-asr table}
\begin{tabular}{|c|c|c|c|}
\hline
\multicolumn{2}{|c|}{}            & PESQ           & STOI           \\ \hline
\multicolumn{2}{|c|}{Noisy}       & 1.563          & 0.650          \\ \hline
\multicolumn{2}{|c|}{SE Baseline} & 1.689          & 0.662          \\ \hline
\multirow{3}{*}{SE-AM}  & Phoneme   & 1.730          & 0.669          \\
                        & BPC(M)  & 1.732          & 0.667          \\ 
                        & BPC(D)  &  \textbf{1.744}         &    \textbf{ 0.674 }    \\ \hline
\multirow{3}{*}{\begin{tabular}[c]{@{}c@{}}SE\\     E2E-ASR\end{tabular}} & Word & 1.688 & 0.667 \\ \cline{2-4} 
                        & Phoneme   & 1.753          & 0.669          \\
                        & BPC(D)  & \textbf{1.803} & \textbf{0.681} \\ \hline
\end{tabular}
\end{table}

% , which corresponds to the most-confusable targets for the recognition model, 
The initial experiments compared the performance of the SE-AM system using BPC(M) and BPC(D) as acoustic units. The results, as shown in Table \ref{tab:am-asr table}, revealed that while BPC(M) provided scores similar to phonemes, BPC(D) outperformed phonemes in terms of both PESQ and STOI scores, indicating that using BPC(D) as the target for the multi-objective training in the SE model is a better choice. Additionally, when using phonemes as the acoustic unit, SE-E2E-ASR outperformed SE-AM in PESQ but had a similar performance in STOI. The word-level SE-E2E-ASR model outperformed the SE baseline in STOI but performed worse in PESQ. Overall, the phoneme-level approaches improved the SE baseline, BPCs-level approaches outperformed the phoneme-level approaches, and the SE-E2E-ASR approaches outperformed the SE-AM approaches.

% Secondly, we compared the SE-AM and SE-E2E-ASR systems using phonemes as acoustic units. The results showed that SE-E2E-ASR outperformed SE-AM in PESQ, but the two systems performed similarly in STOI.}
% % \textcolor{orange}{These results can be explained as follows: The output of E2E-ASR is the consecutive phonemes of the input utterance rather than the tri-phone (inter-phoneme) information provided by the HMM states of the AM. Since multi-objective training} requires E2E-ASR or AM outputs to guide the connected SE model to generate enhanced speech, the resulting phoneme from E2E-ASR would not benefit the SE model and the AM. }
% The word-level SE-E2E-ASR outperformed the transformer-based SE in STOI but was worse in PESQ. The word- and phoneme-level E2E-ASR models can benefit from SE, but they outperform SE-AM only marginally.}

In contrast to the other methods evaluated, the proposed BPC(D)-level SE-E2E-ASR model showed the best performance, with improvements of 0.114 and 0.019 for PESQ and STOI, respectively, compared to the SE baseline. Wilcoxon sign rank tests were performed to measure the performance difference between the BPC(D)-level SE-E2E-ASR and all other methods listed in Table \ref{tab:am-asr table}, including the SE Baseline, three SE-AM strategies, and the Word/Phoneme SE-E2E-ASR approaches. These comparisons were conducted using all 3840 test samples (made up of 192 utterances, 5 unseen noise types, and 4 SNR levels). All the computed p-values were less than the adjusted significance level of 0.01 after applying a Bonferroni correction for multiple tests (0.05/5 = 0.01), providing strong evidence of the superiority of the BPC(D) approach. These results support the hypothesis that incorporating contextual information about the articulation transition between consecutive BPC labels can effectively enhance the quality and intelligibility of processed speech.  

\begin{figure}[t]

\subfigure[PESQ Improvements over the SE baseline]{%
  \includegraphics[clip,width=\columnwidth]{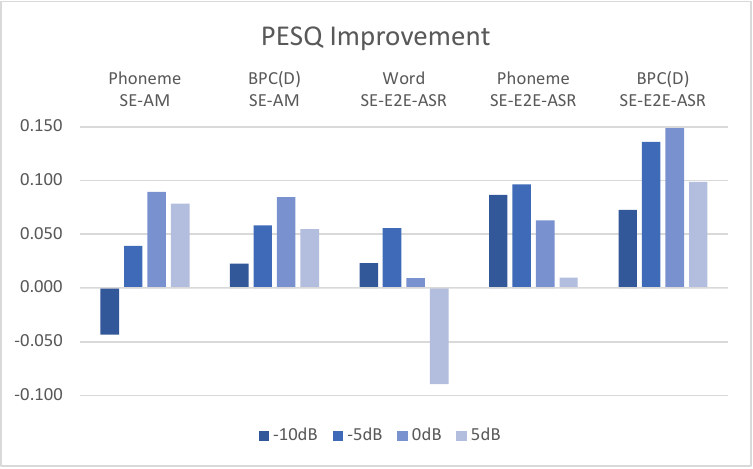}%
  \label{fig:PESQ_Improvement_e2e}
}

\subfigure[STOI Improvements over the SE baseline]{%
  \includegraphics[clip,width=\columnwidth]{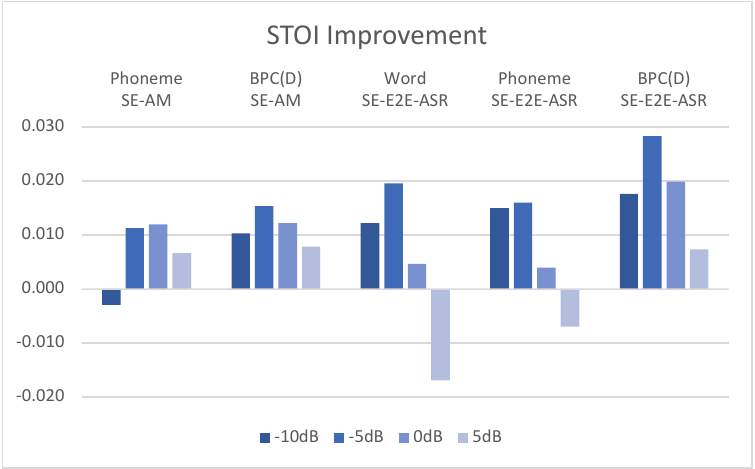}%
  \label{fig:STOI_Improvement_e2e}
}

\caption{Average PESQ and STOI improvements of the two SE-AM and three SE-E2E-ASR systems over the SE baseline on different SNR ranges of the TIMIT corpus.}
\label{fig:E2E Improvement}
\vspace{-4mm}
\end{figure}

Furthermore, we compared these systems in different SNR scenarios, which are shown in Figure \ref{fig:E2E Improvement}. From this figure, we observe that both the phoneme- and BPC-level SE-AMs work well in PESQ and STOI for the three higher SNR cases ($-5$ dB, $0$ dB, and $5$ dB), but perform worse when the SNR is as low as $-10$ dB. This is because it is difficult for the DNN-HMM AM to recognize severely distorted speech and hence it fails to guide the connected SE model. However, the three SE-E2E-ASR systems can enhance the $-10$ dB SNR utterances exceedingly well, showing that E2E-ASR performs better than DNN-HMM AM in promoting SE in low-SNR cases. The BPC-level SE-E2E-ASR provided optimal STOI scores, verifying our hypothesis that contextual broad phonetic information helps in the learning of the SE model.

\subsubsection{Spectrogram for SE-E2E-ASR }
\begin{figure}[htbp]
\centering
\subfigure[Clean]{
\includegraphics[width=4cm]{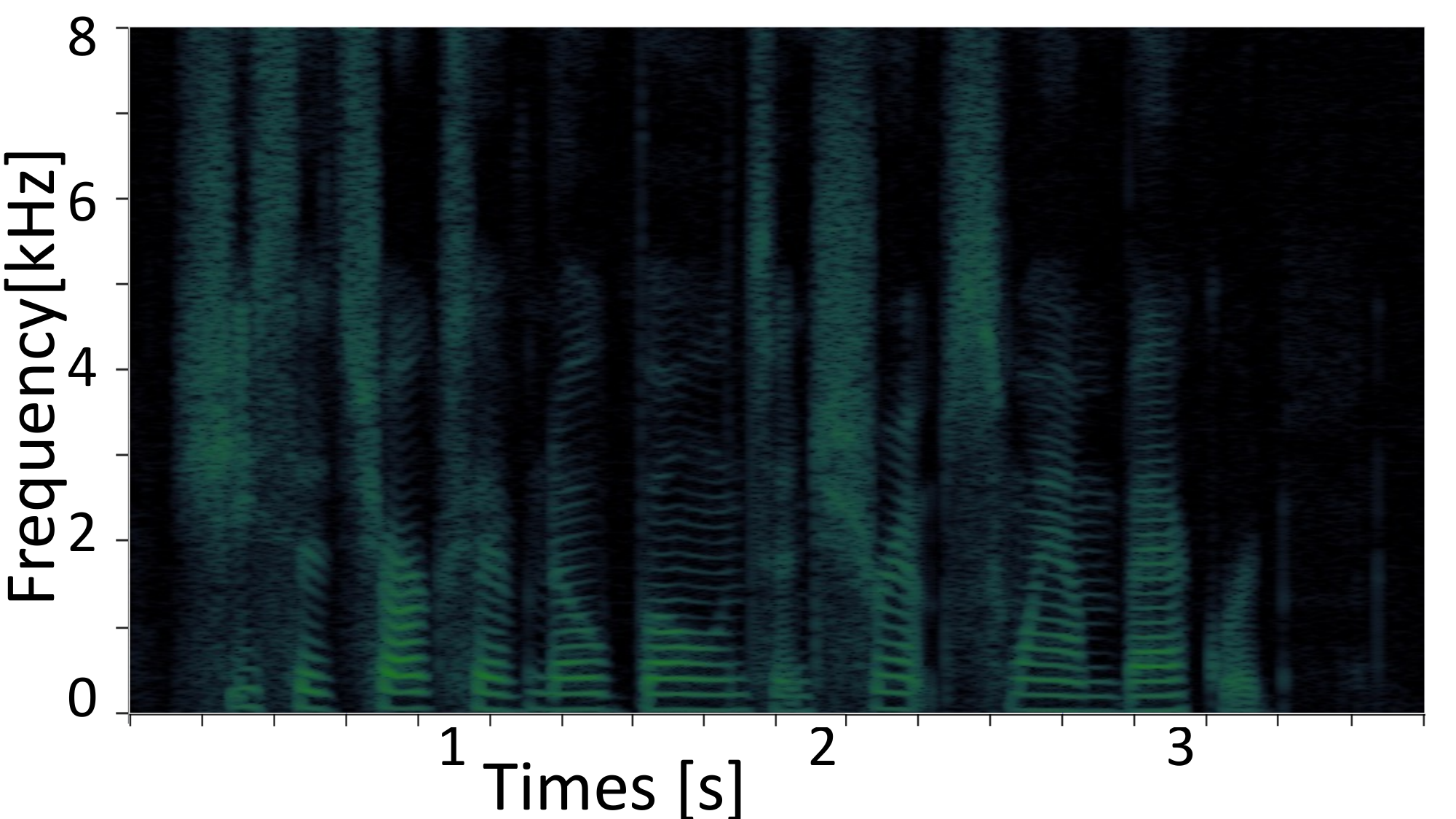}
}
% \quad
\subfigure[Noisy]{
\includegraphics[width=4cm]{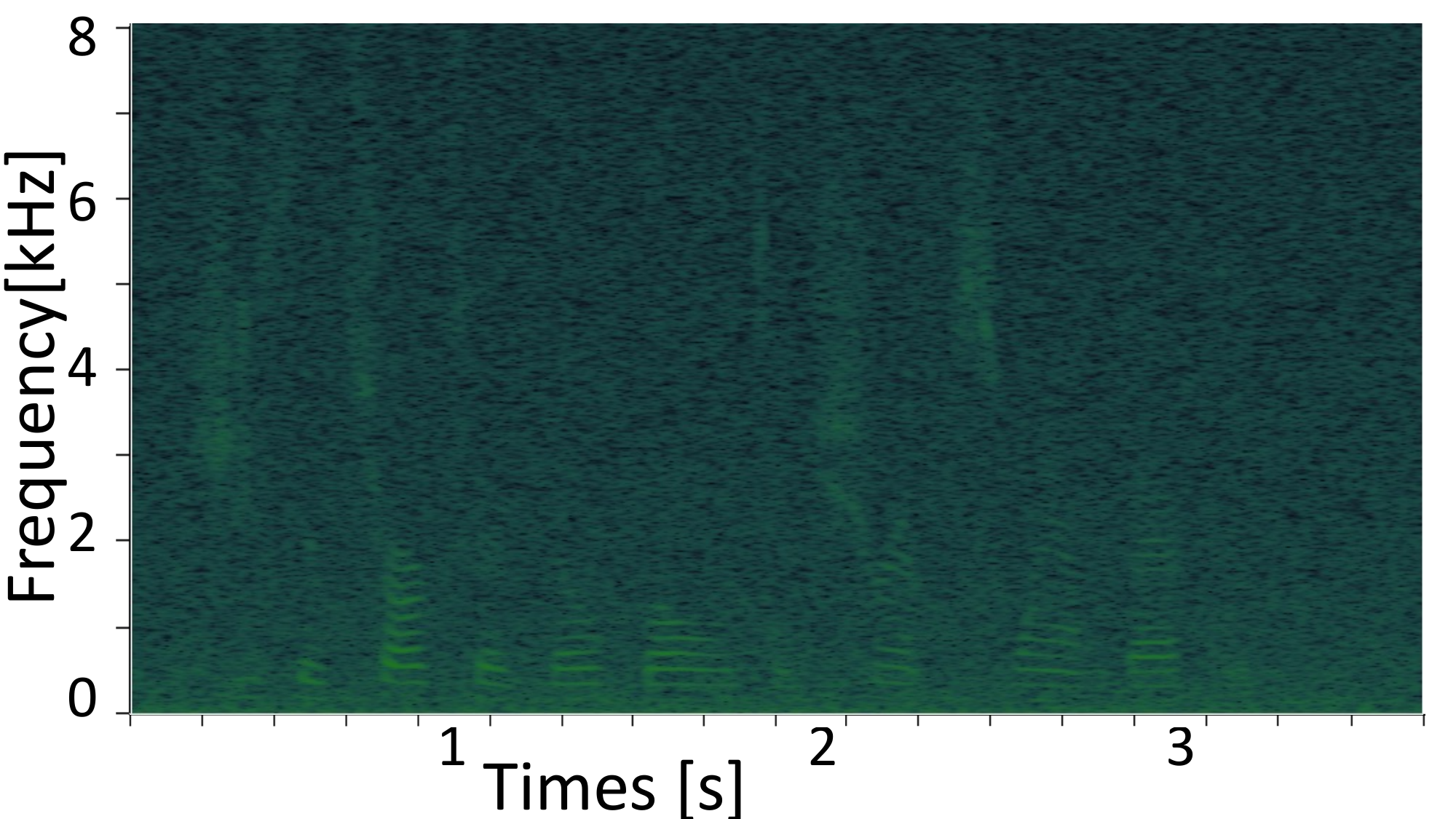}
}
\quad
\subfigure[SE Baseline]{
\includegraphics[width=4cm]{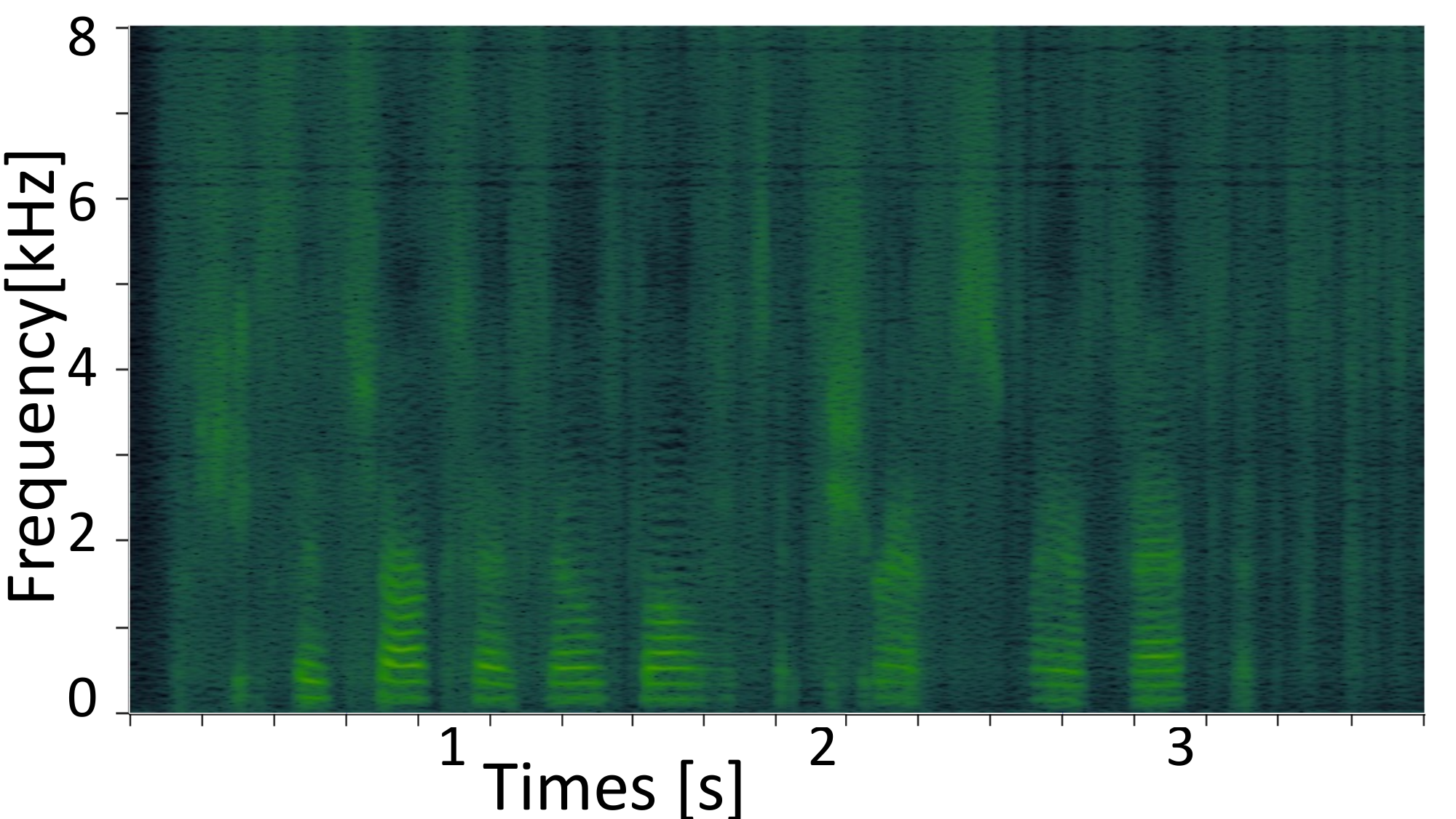}
}
% \quad
\subfigure[BPC(D)]{
\includegraphics[width=4cm]{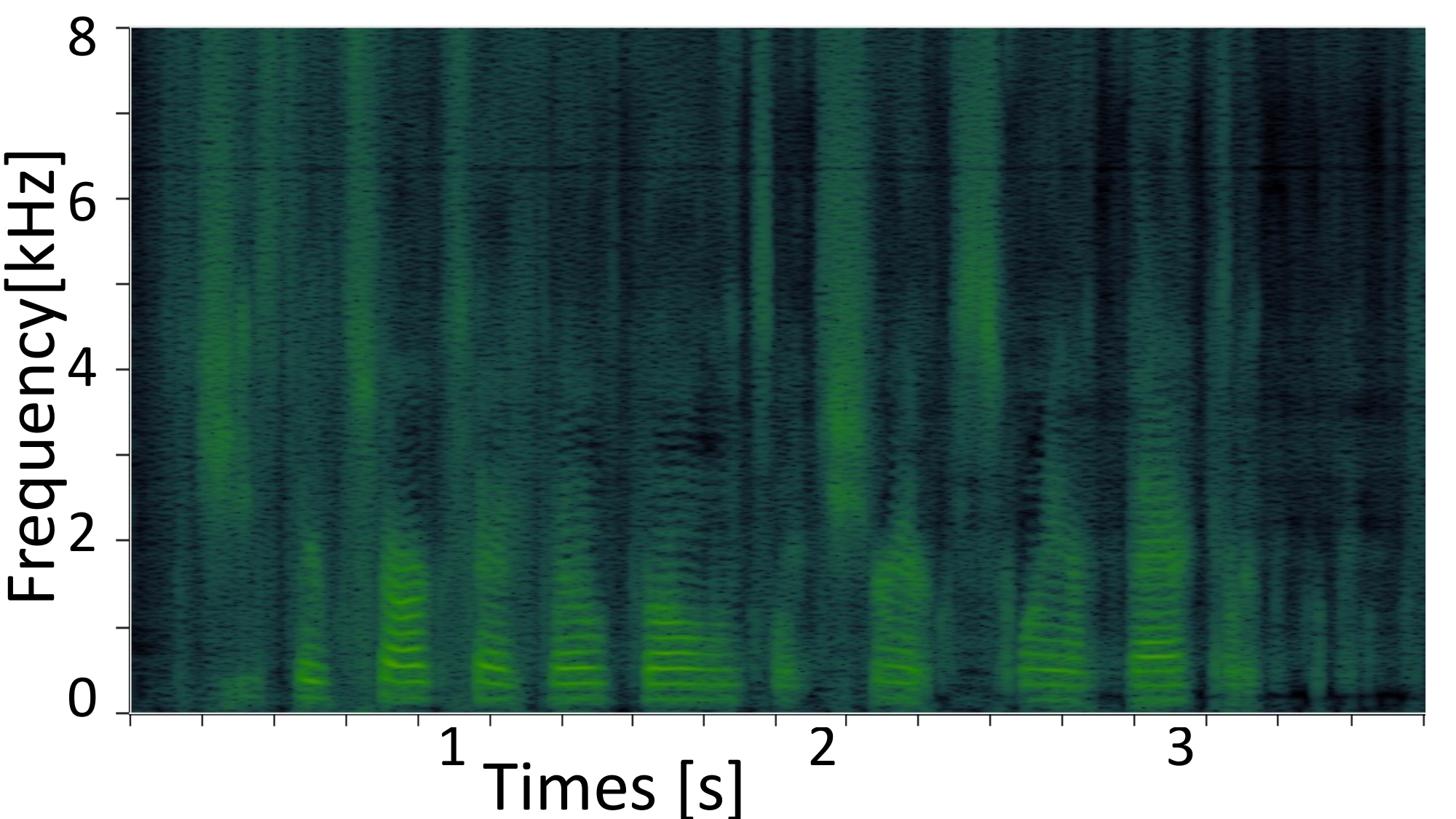}
}

\caption{Spectrogram plots of an utterance at different situations: (a) clean noise-free, (b) noise-corrupted, (c) noise-corrupted and enhanced by the SE baseline, (d) noise-corrupted and enhanced by the BPC(D)-level SE-E2E-ASR.}
\label{fig:timit_spectrogram}
\end{figure}

In addition to the quantitative evaluations, we presented the spectrogram plots for the tested utterance in Figs. \ref{fig:timit_spectrogram} to demonstrate the differences in the SE methods. It shows that the spectrogram of clean speech is severely distorted by noise, while it can be markedly enhanced by the two SE methods. However, there are still moderate residual noises/artifacts in the case of the SE baseline, while they are considerably suppressed by the presented BPC(D)-level SE-E2E-ASR. 

% Fig. \ref{fig:timit_waveform} also shows that the BPC-wise method reduces more noise than the SE baseline, especially in the beginning and ending portions of the utterance. 

\subsubsection{Results for the SE-E2E-ASR with ASR Loss and perceptual loss}
In this subsection, we intend to explore the effects of different losses reported in Eqs. (\ref{eqASR}), (\ref{eqPL}), and (\ref{eqCom}) and thus use the same SE model architecture throughout the experiments. Table \ref{tab:bpc-table} presents the PESQ and STOI scores of the proposed BPC-level SE-E2E-ASR models employing the ASR and perceptual losses, as shown in Fig. \ref{fig:E2E}. Three types of BPC, namely BPC(M), BPC(P), and BPC(D), were individually used as the target for the E2E-ASR model. When using Eqs. (\ref{eqASR}) and (\ref{eqPL}) to train the SE model, we denoted the results as "ASR Loss + L1" and "Perceptual Loss + L1", respectively. For the "Combined Losses" case, the two losses (perceptual loss and ASR loss) from ASR were both used during training as Eq. (\ref{eqCom}). Based on the Table \ref{tab:am-asr table} and \ref{tab:bpc-table}, the following observations were made:

\begin{enumerate}

\item Compared with the results of the phoneme-level SE-AM and SE baseline shown in Table \ref{tab:am-asr table}, almost all the BPC-level SE-E2E-ASR systems achieve better PESQ and STOI scores (except for the BPC(P)-level system with perceptual loss). Integrating SE loss with either ASR loss or perceptual loss (as in Eqs. (2) and (3)) exhibited superior SE performance. Thus, we verified the effectiveness of these contextual and articulatory features in SE.

\item As for the three types of BPCs used in the SE-E2E-ASR with the ASR loss, BPC(P) performed worse than BPC(M) and BPC(D), BPC(D) achieved the optimal PESQ score, and BPC(D) and BPC(M) achieved similar STOI scores. Therefore, the combination of confusion phonemes performed in BPC(D) facilitates the SE module to provide better speech quality, and the clustering methods used in BPC(M) and BPC(D) help to improve speech intelligibility.

\item Regarding the system using the perceptual loss and combined losses, BPC(D) and BPC(M) performed better than BPC(P) in PESQ and STOI metrics, which agreed with the results for the system with the ASR loss. Accordingly, we verified that the performance of the proposed SE-E2E-ASR depends on how we cluster the phonemes.

\item  The experiments with combined losses showed that while the majority of the results outperformed the perceptual loss experiments, they still fell short of the results obtained from the SE E2E-ASR for all types of BPCs. Surprisingly, the PESQ score for the BPC(M) combined losses experiment even outperformed the BPC(M) ASR loss experiment, suggesting that the combination of different loss functions can complement each other and has the potential to further enhance speech quality.

% and the performance of the combined loss is close to the average of ASR loss and perceptual loss experiments. The reason can be that the BPC(D) SE-ASR target already covers the critical information from the ASR model without the need of clean speech.}
% The BPC(D) consistently performs better than other BPCs,
% and BPC(D) has higher PESQ and STOI scores at -10dB in Figure \ref{fig:BPC_Improvement}

\end{enumerate}

\begin{figure}[tp!]

\subfigure[PESQ Improvements over SE baseline]{%
  \includegraphics[clip,width=\columnwidth]{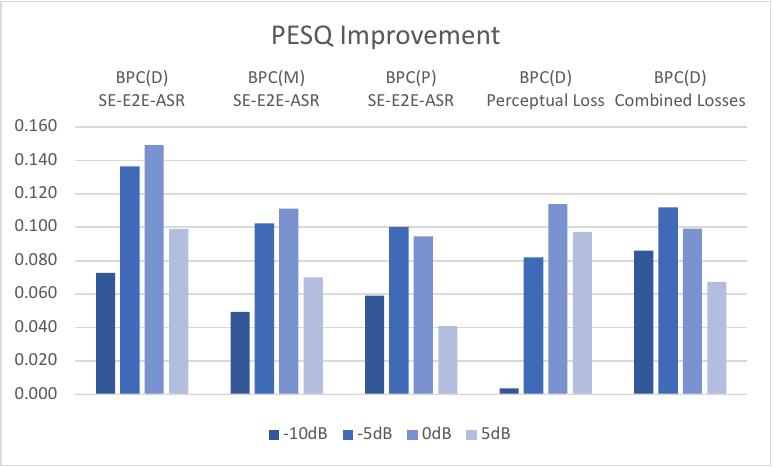}%
  \label{fig:BPC_PESQ_Improvement_TIMIT}
}

\subfigure[STOI Improvements over SE baseline]{%
  \includegraphics[clip,width=\columnwidth]{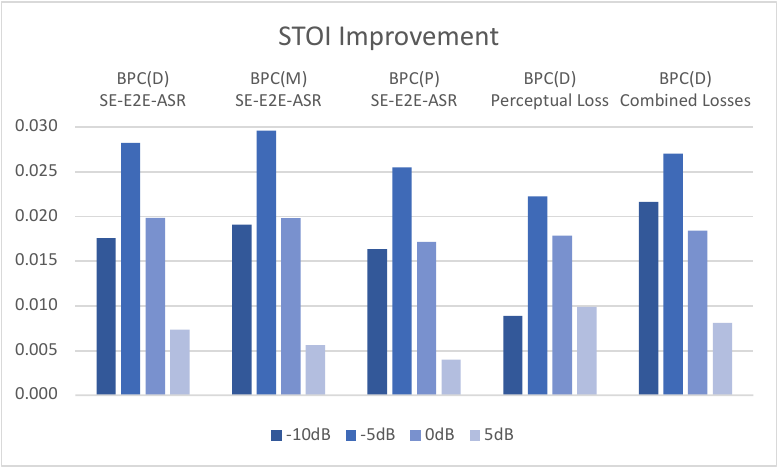}%
  \label{fig:BPC_STOI_Improvement_timit}
}

\caption{Averaged PESQ and STOI improvements of various BPC SE-E2E-ASR systems with ASR and perceptual losses over the SE baseline on different SNR sets for the TIMIT corpus.}
\label{fig:BPC_Improvement}
\vspace{-4mm}
\end{figure}

Even though the systems with perceptual loss achieved lower average PESQ and STOI scores than those with ASR loss, this does not necessarily apply to the individual SNR situation. Figure \ref{fig:BPC_Improvement} shows the PESQ and STOI improvements over the SE baseline for several BPC-level systems at the four SNRs. From this figure, we observe that the three systems with ASR loss exhibit similar trends of improvement with the different SNRs, whereas the BPC(D)-level system with perceptual loss performs quite well in STOI for the high-SNR case (5 dB), outperforming the three systems with ASR loss. On the other hand, the BPC(D)-level system with combined losses performs the best in both PESQ and STOI for the low-SNR case (-10dB).
Furthermore, Note that the ASR loss and the perceptual loss are two types of losses that serve different purposes. The ASR loss aims to improve the accuracy of ASR results, while the perceptual loss measures the difference between clean and enhanced speech at an intermediate level of the ASR model. The results from Table \ref{tab:bpc-table}  suggest that when training SE models, the ASR loss might offer additional and valuable information that complements the L1 loss.

\begin{table}[t!]
\centering
\caption{Average PESQ and STOI scores for the same BPC-level SE-E2E-ASR systems with three losses: ASR,perceptual and combined losses. Scores of the SE baseline are listed for comparison.}
\label{tab:bpc-table}
\begin{tabular}{|c|c|c|c|}
\hline
\multicolumn{2}{|c|}{}                 & PESQ  & STOI  \\ \hline
\multicolumn{2}{|c|}{SE Baseline}      & 1.689 & 0.662 \\ \hline
\multirow{3}{*}{\begin{tabular}[c]{@{}c@{}} ASR Loss + L1 (Eq. (\ref{eqASR}))\end{tabular}} & BPC(M) & 1.772 & \textbf{0.681} \\
                              & BPC(P) & 1.762 & 0.678 \\
                              & BPC(D) & \textbf{1.803} & \textbf{0.681} \\ \hline
\multirow{3}{*}{Perceptual Loss + L1 (Eq. (\ref{eqPL}))} & BPC(M) & \textbf{1.764} & \textbf{0.677} \\
                              & BPC(P) & 1.722 & 0.663 \\
                              & BPC(D) & 1.763 & \textbf{0.677} \\ \hline
\multirow{3}{*}{Combined Losses (Eq. (\ref{eqCom}))} & BPC(M) & 1.777 & 0.677 \\
                              & BPC(P) & 1.746 & 0.667 \\ 
                              & BPC(D) & \textbf{1.780} & \textbf{0.681} \\ \hline
\end{tabular}
\end{table}

\subsection{Overall discussion of the TIMIT experiments}

\subsubsection{Misclassification of phonemes causes poor feedback}
\label{sec:timit misclass}
From the experiments on the TIMIT dataset, we observed that misclassification of phonemes by the ASR system can lead to poor feedback, as discussed below:

\begin{enumerate}
\renewcommand{\labelenumi}{\alph{enumi}.}
\item Our experiments shows that BPC-level objectives outperformed phoneme-level objectives, suggesting that distinguishing between confusable phonemes may not be as helpful as correctly classifying groups of phonemes. When the SE model learns to generate speech that overly emphasizes the difference between similar phonemes, the generated speech may not necessarily be an improvement.

\item We also found that correct objective feedback from the ASR loss performs better than the soft objective from the perceptual loss. This indicates that misclassification results from clean speech reduce the improvement from the ASR feedback, highlighting the importance of accurate feedback for effective model training.

\item BPC(D) performs the best in almost all the experiments, indicating that combining the most-confusable targets is the most helpful for the additional objective. On the other hand, the place of articulation is not as critical as the manner of articulation for the shape of the audio waveform \cite{ladefoged2014course}, meaning that the phonemes in the same group of BPC(P) are not confusable and lead to the worst performance among all the BPCs.

\end{enumerate}

Based on the above observations, we conclude that objectives with misclassification of phonemes by the ASR system may lead to inadequate feedback for SE models. Although the experiments were conducted on a relatively small set of training data (3,696 clean utterances), the performance improvement of low-resource training conditions is still valuable for practical application. However, it is worth noting that the advantages of the knowledge-based approach (BPC(M) and BPC(P)) may be reduced as the amount of data increases.

\subsubsection{Contextual acoustic feedback from the E2E-ASR model}
% Discuss the SE-AM, SE-E2E-ASR-word, SE-E2E-ASR-acoustic, and perceptual loss methods, with other literature
Most previous studies that apply feedback from the ASR objective for SE use losses from AM, which provides frame-wise feedback \cite{subramanian2019speech}, or contextual E2E-ASR with word-level objective feedback \cite{bagchi2018spectral, plantinga2021perceptual}. In contrast, our approach applies phoneme-level E2E-ASR feedback for the SE system. This approach has benefits in that the ASR model learns to predict phonemes as a sequence instead of individually for each time segment. The benefits of this approach are listed below:

\begin{enumerate}
\renewcommand{\labelenumi}{\alph{enumi}.}
\item Compared to AM feedback like \cite{bagchi2018spectral, plantinga2021perceptual}, phoneme/BPCs-level E2E-ASR feedback can guide the SE model with the level of the whole utterance instead of individual time segments. One of the advantages of using phoneme-level E2E-ASR feedback for SE is that it allows for better modeling of the temporal relationships between speech features and phonemes. In traditional ASR systems, phonemes are typically modeled using hidden Markov models (HMMs), which do not take into account the temporal structure of the speech signals. However, in phoneme-level E2E-ASR, the ASR model learns to predict phonemes as a sequence, which allows for better modeling of the dynamic relationships between speech features and phonemes. The consistent results of Phoneme/BPC(M)/BPC(D)-level E2E-ASR feedback outperform corresponding AM feedbacks in Table \ref{tab:am-asr table} and Table \ref{tab:bpc-table}, supporting this statement.

\item Another advantage of using phoneme/BPCs-level E2E-ASR feedback for SE is that it provides more direct and informative feedback to the SE model compared with word-level E2E-ASR feedback like \cite{subramanian2019speech}. As we mentioned earlier, phonemes are closer to the audio features compared to words, which makes the phoneme-level feedback more relevant and useful for guiding the SE model. Moreover, phoneme-level feedback is more fine-grained than word-level feedback, which allows for better differentiation of the different phonemes and their acoustic characteristics. The results of Word-level E2E-ASR feedback perform the worst among the Word/Phoneme/BPC(D)-levels in Table \ref{tab:am-asr table}, supporting this statement.
\end{enumerate}

 These observations show that incorporating contextual broad phonetic information to learn the SE model, as in BPC-level SE-E2E-ASR, is most helpful in reconstructing the original clean signal and removing the interference. It's worth noting that while phoneme/BPCs-level E2E-ASR feedback has several advantages for SE, it also has some potential limitations. For example, phoneme/BPCs-level ASR feedback may require transcription training labels compared to other types of feedback such as perceptual loss. This can be a limitation in low-resource settings where obtaining large amounts of labeled data is challenging. Additionally, the phoneme-level feedback may not be as effective for languages with complex phonetic systems. However, the consistent improvement of the Phoneme/BPC(M)/BPC(D)-level E2E-ASR feedback over the AM feedback and the Word-level E2E-ASR feedback in our experiments suggests that phoneme/BPCs-level E2E-ASR feedback is a promising approach for improving the performance of SE systems.

% \subsubsection{The size of model and dataset}}

\subsection{Experiments on the TMHINT dataset}
International Phonetic Alphabet (IPA) presents the phones used in all languages. Therefore, articulation feature classification methods using BPCs can also be applied to other languages. For the TMHINT corpus, we transferred the Chinese characters into IPA phone sequences and categorized these phones into BPC clusters.

For the training set, we used 10,000 noisy-clean paired training utterances. The paired training set contained 1,200 clean utterances with an average duration of 3.5 seconds and their noisy mixed speech using 104 noises with multiple noise sources \cite{hu2004100} (at 31 SNR levels from -10 to 20 dB). For the test set, 640 utterances were mixed with seven unseen noise types at 14 SNR levels (ranging from -10 to 10 dB). The training set included three male and three female speakers, and the testing set contained one male and one female. The SE-E2E-ASR experiments used various ASR labels, such as phoneme, BPC(M), and BPC(P). Furthermore, we evaluate the ability of the clustering approach in English used in BPC(D) to generalize to other languages. Specifically, we apply the BPC(D) clusters trained on English data to a denoising experiment on TMHINT by mapping the Chinese phonemes to their corresponding IPA symbols. We made some adjustments to the original BPC(D) clusters by removing redundant phonemes and grouping new phonemes based on their manner of articulation, resulting in a total of nine groups, as in the English experiment. This evaluation allows us to test the generalizability of the data-driven approach across different languages and to assess whether the phoneme clusters learned from one language can be applied to another language. The transformer was set as the baseline SE model and learned jointly with the connected ASR model with extra BPC semantic information. A BPC(M)-ASR classification model trained with noisy speech was also examined. For the SE-only task, the same perceptual loss in Eq. (\ref{eqPL})  used in the earlier TIMIT experiment presented the informative semantic features in the ASR model. To test the SE-E2E-ASR methods in the experiments for noisy-reverberant utterances, we selected BPC(M) as the recognition unit.

% \begin{figure}[htbp]
% \centering
% \subfigure[Clean]{
% \includegraphics[width=4cm]{figures/Clean_wav.pdf}
% }
% % \quad
% \subfigure[Noisy]{
% \includegraphics[width=4cm]{figures/Noisy_wav.pdf}
% }
% \quad
% \subfigure[Baseline]{
% \includegraphics[width=4cm]{figures/Baseline_wav.pdf}
% }
% % \quad
% \subfigure[BPC(M)]{
% \includegraphics[width=4cm]{figures/BPC_wav.pdf}
% }

% \caption{Waveform plots of an utterance at different situations: (a) clean noise-free, (b) noise-corrupted, (c) noise-corrupted and enhanced by the SE baseline , (d) noise-corrupted and enhanced by the BPC(M)-level SE-E2E-ASR.}
% \label{fig:timit_waveform}
% \end{figure}

\begin{table}[t]
\centering
\caption{Averaged PESQ and STOI scores for SE-E2E-ASR system on TMHINT corpus.}
\label{tab:TMHINT_Result}
\begin{tabular}{|cc|cc|}
\hline
 &  & PESQ & STOI \\ \hline
\multicolumn{2}{|c|}{Noisy} & 1.572 & 0.684 \\ \hline
\multicolumn{2}{|c|}{SE Baseline} & 2.029 & 0.725 \\ \hline
\multicolumn{1}{|c|}{\multirow{5}{*}{\begin{tabular}[c]{@{}c@{}}SE \\ E2E-ASR\end{tabular}}} & Phoneme & 2.012 & 0.724 \\
 \cline{2-4} 
\multicolumn{1}{|c|}{} & BPC(M) & \textbf{2.068} & \textbf{0.729} \\
\multicolumn{1}{|c|}{} & BPC(P) & 2.060 & 0.728 \\
\multicolumn{1}{|c|}{} & BPC(D) & 2.047 & \textbf{0.729} \\
 \cline{2-4} 
\multicolumn{1}{|c|}{} & BPC(M)-MT & \textbf{2.066} & \textbf{0.731} \\ \hline
\multicolumn{1}{|c|}{Perceptual Loss} & BPC(M) & \multicolumn{1}{l}{2.034} & \multicolumn{1}{l|}{0.726} \\ \hline
\end{tabular}
\end{table}

\subsubsection{Results for Speech Denoising}
The model structure used here was similar to that described in Section \ref{sssec:timit_exp}. As the value of the ASR loss was larger than the SE loss in this task, we lowered the parameter $\alpha$ in Eq. (\ref{eqASR}) to 0.0001 to equally weight the two losses in the total loss function. The resulting PESQ and STOI scores are presented in Table \ref{tab:TMHINT_Result}. From this table, we first observe that the BPC-level SE-E2E-ASR model can improve both the PESQ and STOI scores compared to the SE baseline, while the mono-phoneme SE-E2E-ASR model compromises the SE improvement. These results differ from those obtained for the TIMIT task described in Section \ref{sssec:timit_exp}. This might be because Mandarin Chinese is a tonal language, where the classification of respective phonemes may be less helpful for the SE model.  Second, when the English data-driven cluster BPC(D) is applied to the TMHINT corpus, it produces the lowest PESQ score among the BPC-level SE-E2E-ASR approaches, despite its superior performance compared to other English clusters. Our investigation in \ref{sec:timit misclass} suggests that the inaccurate classification of phonemes can lead to unsatisfactory results. Considering the distinctive acoustic properties of Chinese, it is reasonable to assume that BPC(D) needs further customization, including the incorporation of tonal features, for Chinese corpora. To validate this hypothesis, future studies can explore various phoneme groups specifically designed for the Chinese language. Comparatively, using BPC(M) as the acoustic unit for SE-E2E-ASR resulted in the best performance among all selections, including monophonic, BPC(P), and BPC(D). Since BPC(M) (and BPC(P)) is designed based on the property of IPA-level phones, the cluster is identical to the BPC(M) we use in English and could potentially be used for cross-language SE training in the future.

In addition to the experiments, wherein the E2E-ASR model was trained with clean, noise-free utterances, we used noisy utterances to train the E2E-ASR model and then conducted the respective SE-E2E-ASR experiments. We randomly selected the noise signals within 104 different noise types mixed with 1,200 utterances as the training set to train the CTC/attention E2E-ASR model. The subsequent SE-E2E-ASR experiments with the multi-condition trained E2E-ASR model adopted the same training configuration as those with the clean E2E-ASR model mentioned above, with BPC(M) as the acoustic unit. The obtained PESQ and STOI scores, which are listed in Table \ref{tab:TMHINT_Result} with the label ``BPC(M)-MT,” are quite close to those of the clean BPC(M)-level E2E-ASR model. These results clearly show that whether the ASR model is trained by clean or noisy utterances does not considerably influence the average SE performance of SE-E2E-ASR if BPC(M) is used.

\begin{figure}[tp!]

\subfigure[PESQ Improvements over the SE baseline]{%
  \includegraphics[clip,width=\columnwidth]{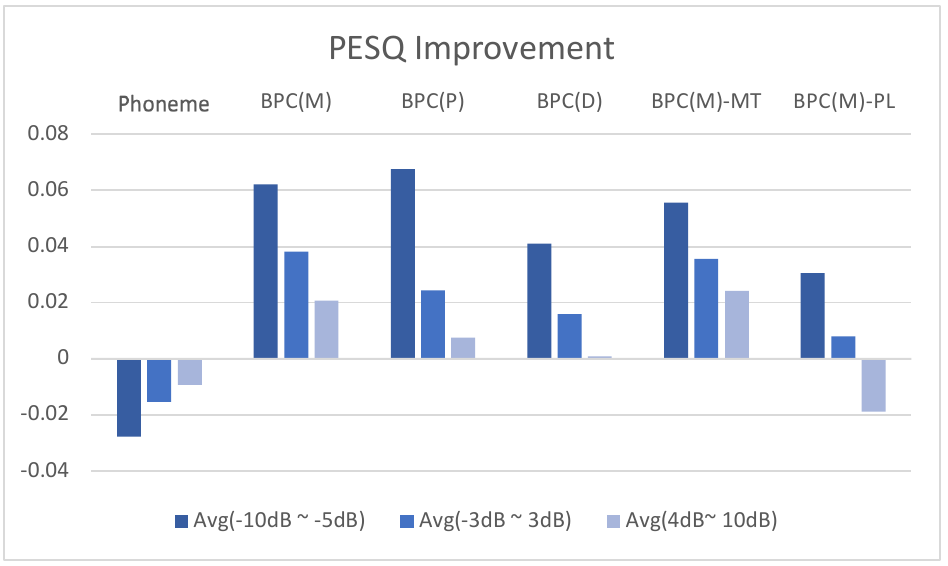}%
  \label{fig:PESQ_Improvement_transformer}
}

\subfigure[STOI Improvements over the SE baseline]{%
  \includegraphics[clip,width=\columnwidth]{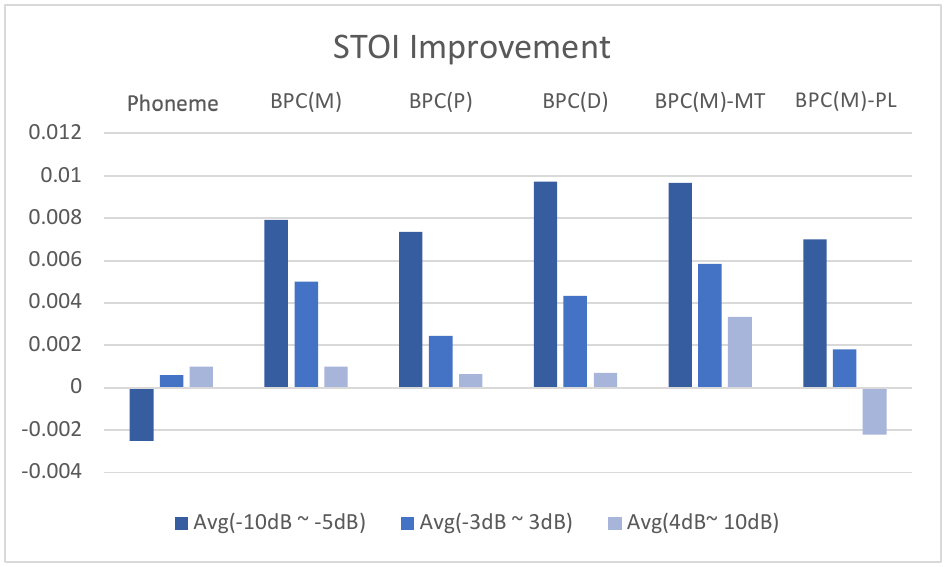}%
  \label{fig:STOI_Improvement}
}

\caption{The PESQ and STOI improvements of SE-E2E-ASR system over the SE baseline averaged on different SNR sets for the TMHINT corpus.}
\label{fig:Improvement}
\vspace{-4mm}
\end{figure}

Figure \ref{fig:Improvement} shows the PESQ and STOI improvements of different SE-E2E-ASR systems over the SE baseline at different SNR sets (-10-5 dB, -3-3 dB, and 4-10 dB). As shown in the figure, almost all systems outperform the SE baseline, except for the phoneme-level system and the perceptual loss system (at high SNRs). These results demonstrated the effectiveness of articulatory features of BPCs for SE particularly for exceedingly noisy (low-SNR) situations. By contrast, the phoneme-level E2E-ASR may not benefit the connected SE at low SNRs probably due to its poor recognition accuracy.

\begin{figure}[htbp]
\centering
\subfigure[Clean]{
\includegraphics[width=4cm]{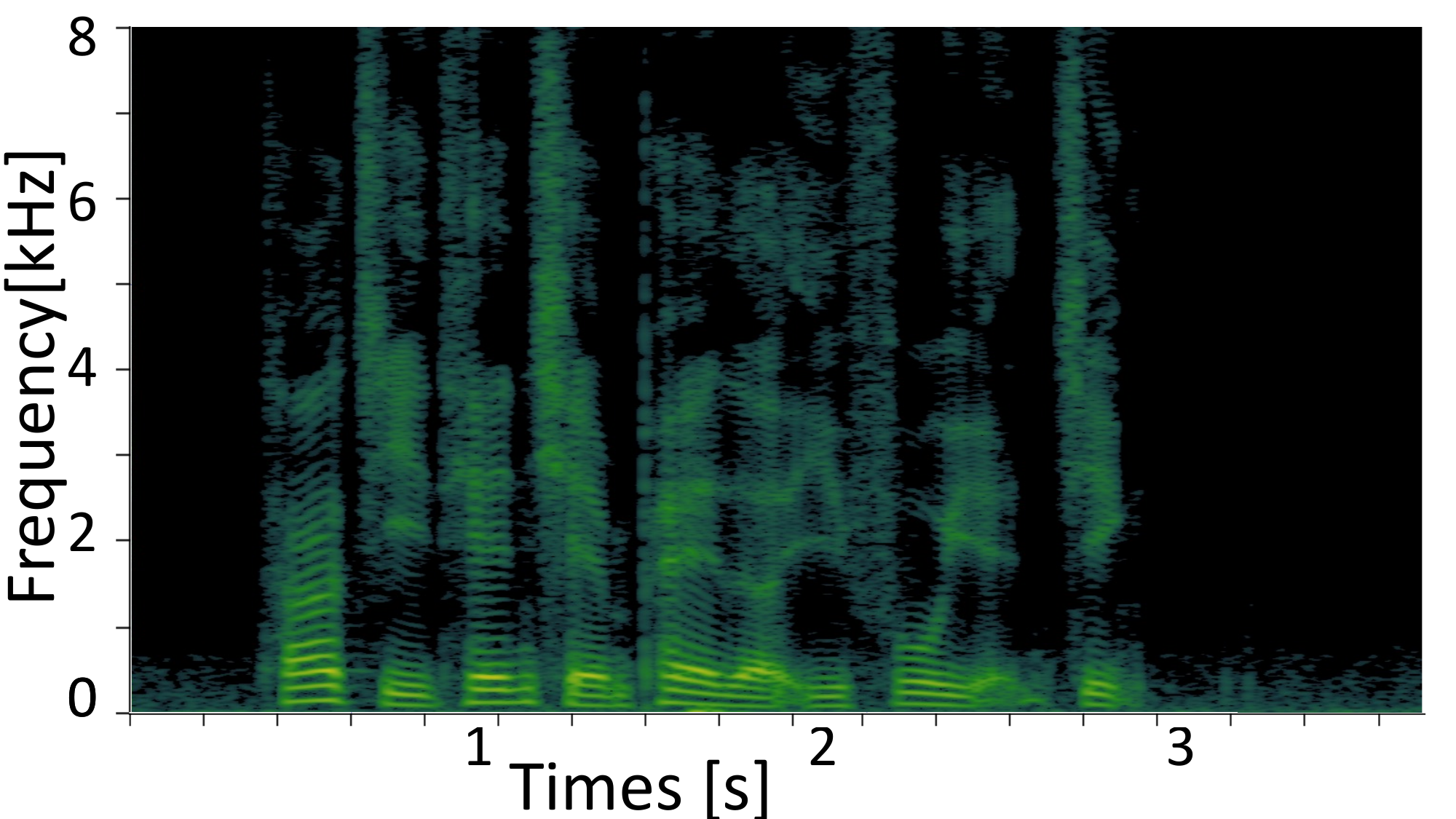}
}
% \quad
\subfigure[Noisy]{
\includegraphics[width=4cm]{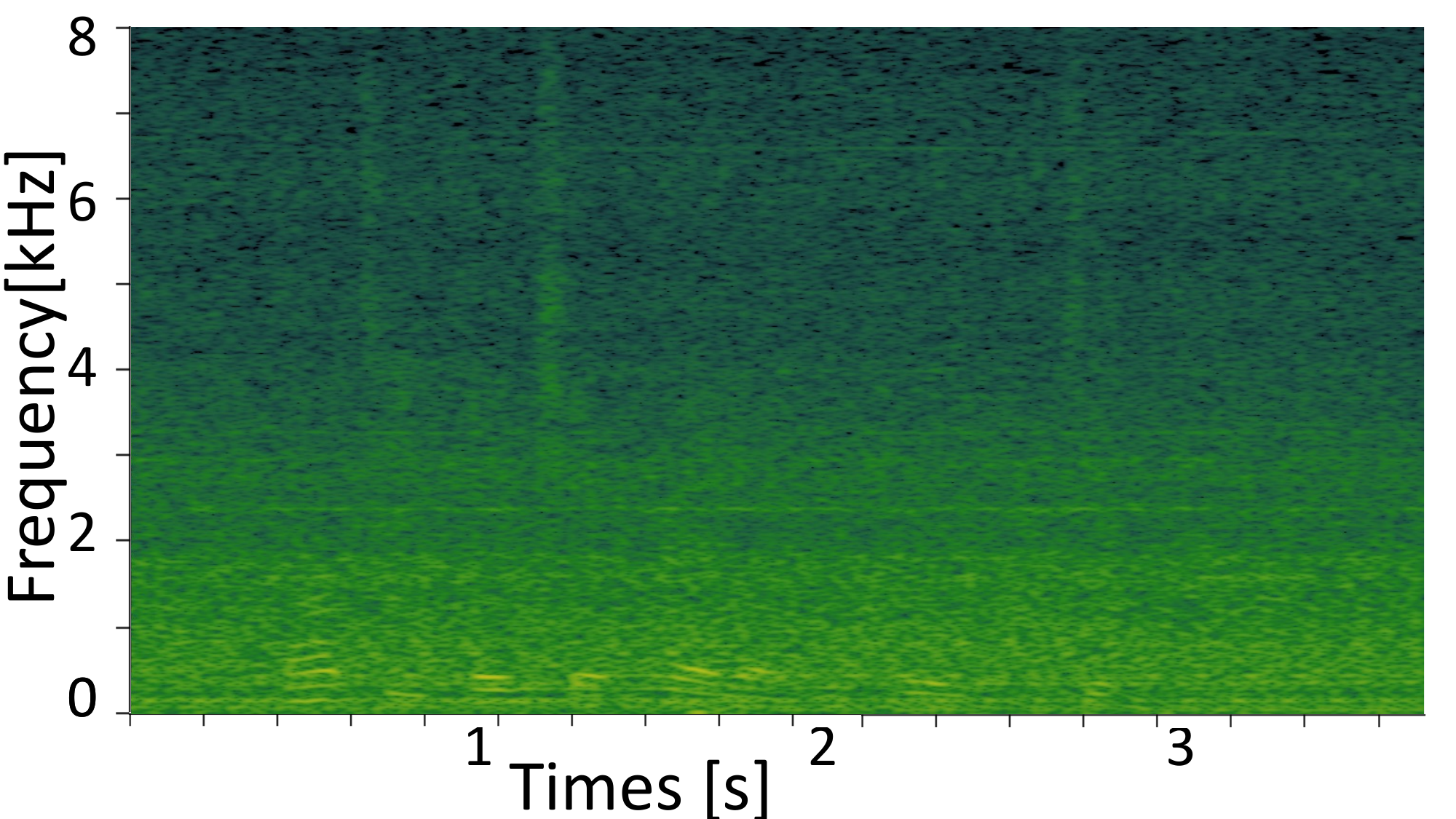}
}
\quad
\subfigure[Baseline]{
\includegraphics[width=4cm]{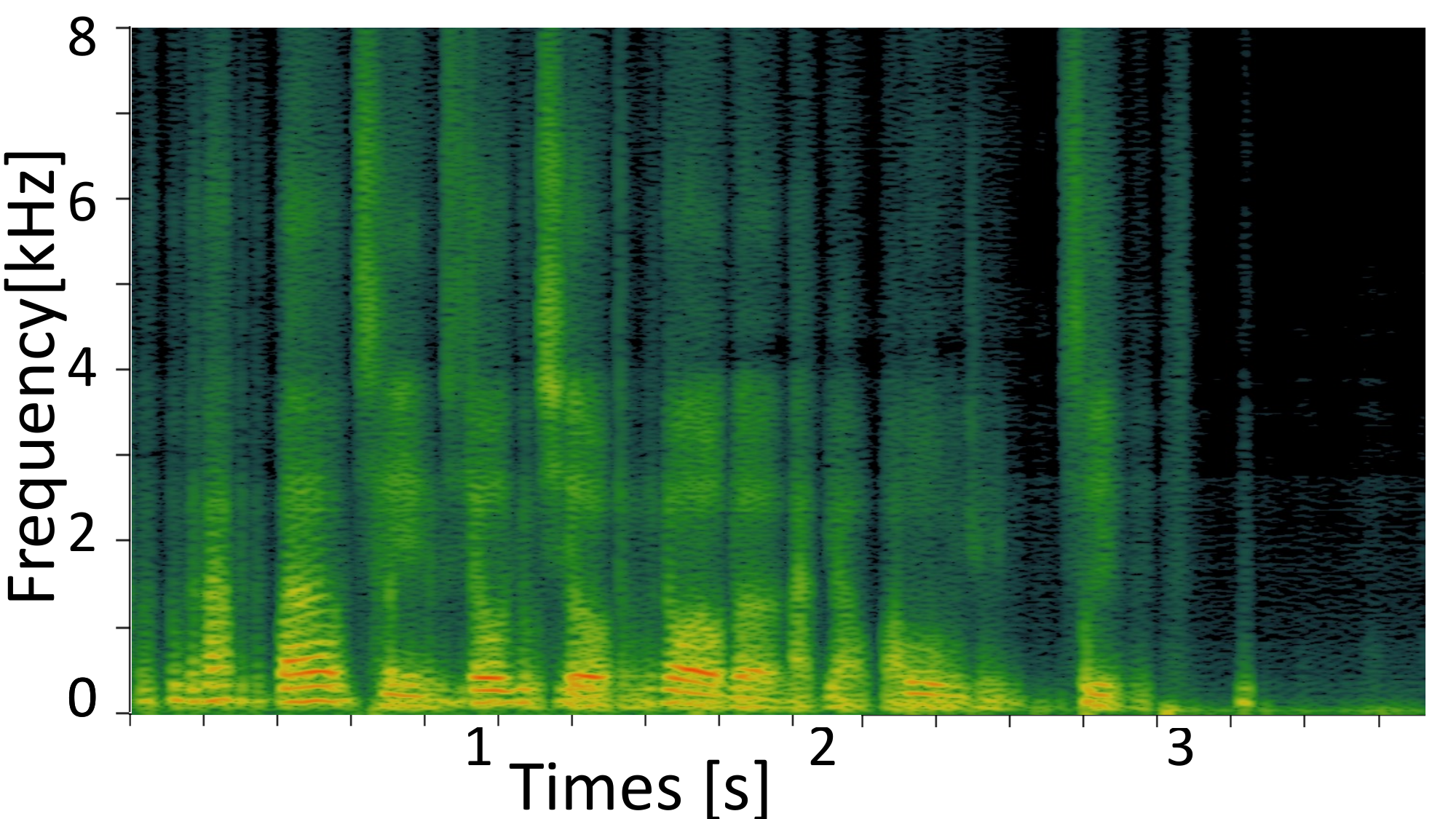}
}
% \quad
\subfigure[BPC(M)]{
\includegraphics[width=4cm]{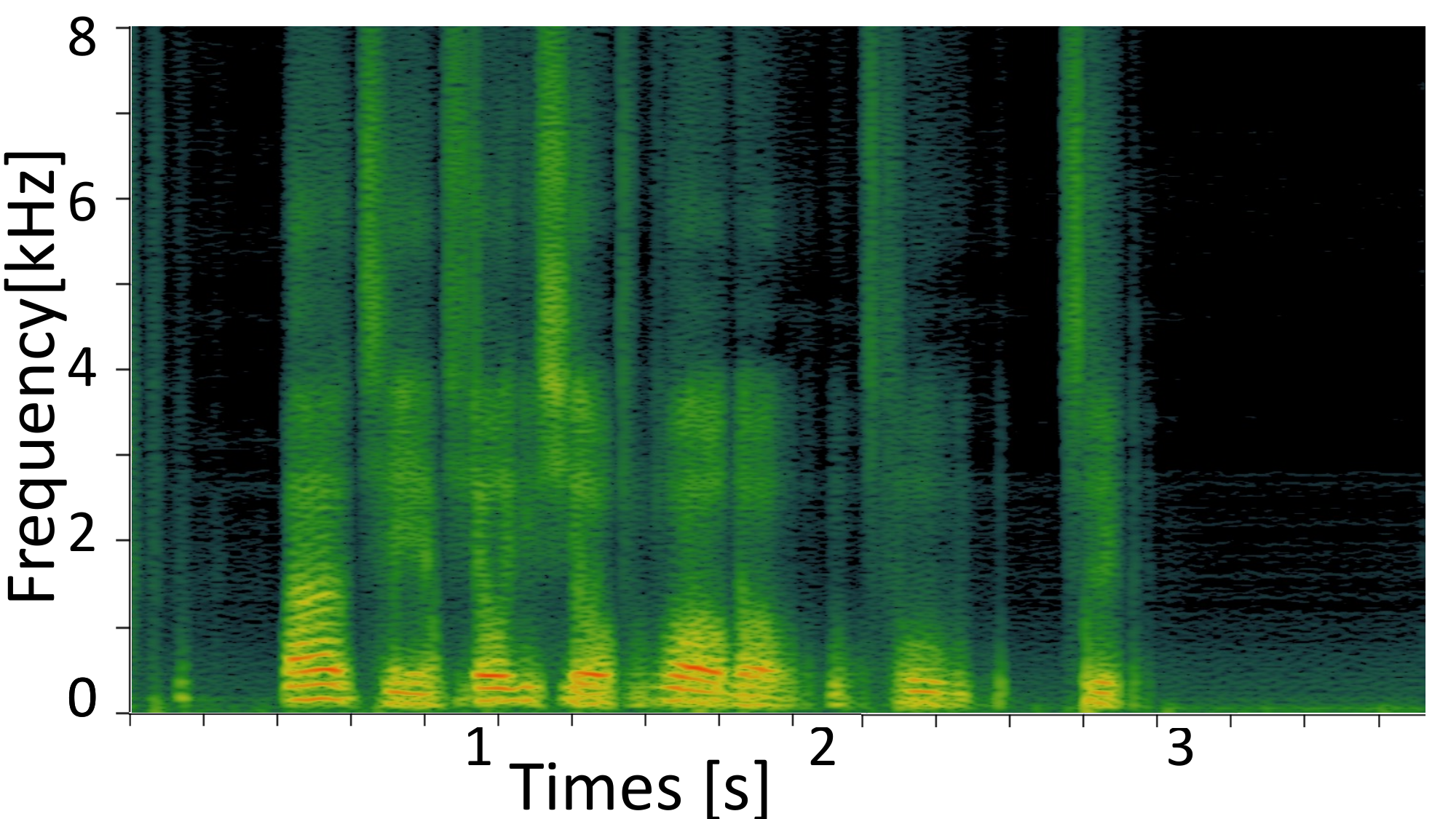}
}

\caption{Spectrogram plots of an utterance at different situations: (a) clean noise-free, (b) noise-corrupted, (c) noise-corrupted and enhanced by the SE baseline, (d) noise-corrupted and enhanced by the BPC(M)-level SE-E2E-ASR.}
\label{fig:tmhint_spectrogram}
\end{figure}

Additionally, in Figure \ref{fig:tmhint_spectrogram}, we display the spectrograms of the clean speech, its noisy counterpart, and their enhanced versions at an engine noise SNR of -5dB for qualitative comparison. It is evident from the figure that the SE baseline fails to entirely remove the noise in non-speech regions, whereas the proposed BPC(M)-level SE-E2E-ASR model better suppresses the noise in these areas, resulting in a spectrogram that more closely resembles clean speech. This finding reaffirms that the contextual information of the BPCs enhances SE performance.

% Furthermore, in Fig. \ref{fig:tmhint_spectrogram}, we present the spectrograms of the clean utterance, noisy counterparts, and their enhanced versions at -5dB SNRs engine noise} for qualitative comparison. From this figure, we observe that the SE baseline does not completely remove the noise in the non-speech regions, whereas, The noise in these regions is better suppressed, resulting in a spectrogram that appears to be more similar to clean speech for the proposed BPC(M)-level SE-E2E-ASR case}. Accordingly, we reconfirmed that the contextual information of the BPCs boosts SE performance. 
% We highlighted three non-speech regions in the clean utterance and used them to compare the enhanced counterparts from the SE baseline and BPC(M)-level SE-E2E-ASR. 

% \begin{figure}
%  \centering
%  \includegraphics[width=\linewidth]{./figures/TMHINT_spec.pdf}
%  \caption{The spectrograms of an utterance and its enhanced versions from the TMHINT corpus at different SNRs.} 
%  \label{fig:TMHINT spec}
% \end{figure}

\subsubsection{Listening Test}

To further evaluate the effectiveness of the proposed approach, a listening test was conducted for the TMHINT experiments.The test set included two challenging noise types - engine and street noises - with two different SNR levels (-5 and 5 dB). The four processing approaches - Baseline, BPC(M) with perceptual loss (with loss of Eq. (2)), Phoneme, and BPC(M) (with loss of Eq. (1))- were tested with each noise type and SNR level as a total of 16 conditions: 2 SNR level × 2 noise types × 4 processing approaches, each containing ten randomly selected sentences. The order of the conditions was also randomized. The subjective quality of the enhanced speech utterances was evaluated using mean opinion score (MOS) tests, with 17 subjects asked to judge the quality of the audio for signal distortion (SIG), background intrusiveness (BAK), and overall quality (OVL) using a five-point scale (1: Bad, 2: Poor, 3: Fair, 4: Good, 5: Excellent) \cite{hu2007evaluation}.

\begin{figure}[t]
  \centering
  \subfigure[scores of signal distortion (SIG)]{\includegraphics[width=0.15\textwidth]{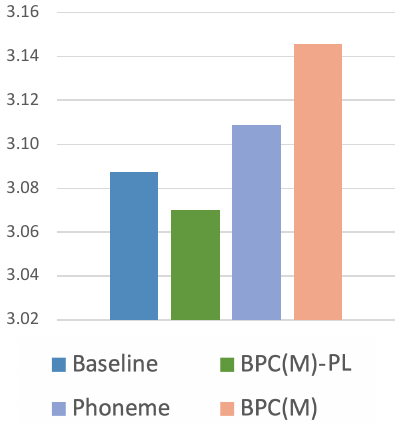} \label{fig:listening test sig}}
  \subfigure[scores of background intrusiveness (BAK)]{\includegraphics[width=0.15\textwidth]{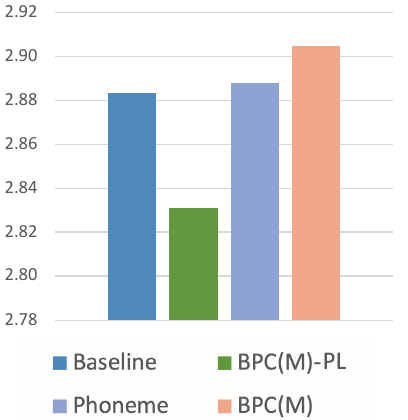} \label{fig:listening test bak}}
  \subfigure[scores of overall quality (OVL)]{\includegraphics[width=0.15\textwidth]{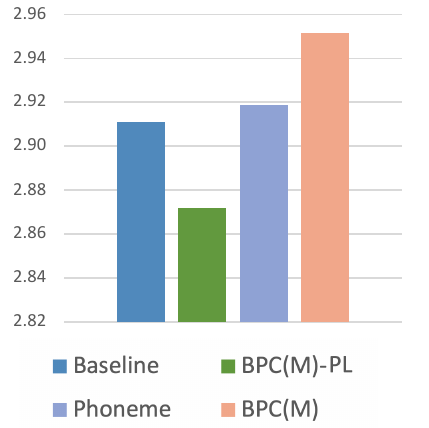} \label{fig:listening test ovl}}
  \caption{Listening test results in terms of SIG, BAK, and OVL scores. The "Phoneme" and "BPC(M)" are the approaches with E2E-ASR objectives, and the "BPC(M)-PL" is the approach with the perceptual loss.}
  \label{fig:listening test}
\end{figure}

In the SIG test, the subjects were asked to rate the natural level of the speech signals after listening to an enhanced speech utterance processed by the four different methods. A higher score indicates that the speech signals are more natural. The SIG results are shown in Figure \ref{fig:listening test sig}. We found that although the perceptual loss approach (BPC(M)-PL) performs worse than the results from the baseline, the ASR objectives (Phoneme and BPC(M)) improve the scores, and BPC(M) provides the best result. This shows that the E2E-ASR objective alleviates the distortion of the recovered signal and provides better speech quality.
For the BAK test, the subjects were asked to judge the level of noise artifact perceived after listening to an utterance, and a higher score indicated a lower level of noise artifact perceived. In \ref{fig:listening test bak}, BPC(M) was the only method that outperformed the baseline in BAK scores, while Phoneme performed close to the baseline and the BPC(M)-PL approach decreased the score. Finally, for the overall quality (OVL) in \ref{fig:listening test ovl}, while the BPC(M)-PL still compromises the results, the results of the Phoneme approach are slightly better than the results from the baseline, and the BPC(M) approach performs the best in the overall quality. The results indicate that while the perceptual loss approach did not perform well in subjective evaluations, both Phoneme-level and BPC-level E2E-ASR objectives yielded less distorted outcomes, and BPC targets appeared to be more effective in eliminating noise and providing better speech quality. To supplement our analysis, we also conducted the Wilcoxon sign rank tests to compare the four approaches in terms of overall quality (OVL), background intrusiveness (BAK), and signal distortion (SIG) scores. The test results show that in terms of OVL and BAK scores, there is a statistically significant improvement when comparing the BPC(M) method with the Baseline and BPC(M)-PL approaches (p $<$ 0.05). However, for SIG scores and the Phoneme approach, no significant differences were found. One possible reason for the lack of significant difference in SIG scores and the Phoneme approach might be due to a high variation among the listeners' responses, combined with a relatively small sample size. These factors make it more challenging to reach statistically significant results.

\subsubsection{ASR Results}

We compared the performance of different SE approaches on automatic speech recognition (ASR) using Google Speech-to-Text \cite{zhang2017speech} to compute word error rates (WER) for baseline and SE-E2E-ASR systems. We conducted experiments with four real-world noise types - engine, babble, street, and three-talker - to simulate practical environments. Table \ref{tab:wer} summarizes our findings. Specifically, we evaluated the WER of speech enhanced using transformer baseline, BPC(M)-PL, Phoneme, and BPC(M)  objectives. Our results indicate that all Phoneme/BPC-level objectives (i.e., Phoneme, BPC(M), and BPC(M)-PL) resulted in a reduced WER compared to the SE baseline. Notebally, the BPC-based enhancement methods (BPC(M) and BPC(M)-PL) surpassed the Phoneme-level SE-E2E-ASR approach. Among all, BPC(M)-PL yielded the best results across all signal-to-noise ratio (SNR) conditions.

Our experiments demonstrate that BPC-based enhancement approaches are promising pre-processing modules for speech recognition in practical settings. Our findings indicate that the perceptual loss approach, despite some compromises in subjective evaluation, yields more accurate recognition results than the Phoneme and baseline approaches. Meanwhile, the Phoneme/BPC-level E2E-ASR objective approaches strike a balance between subjective evaluation and recognition results, achieving improved performance across all evaluation metrics. Specifically, the BPC(M)-level E2E-ASR objective approach outperformed all the other approaches in all experiments.
% Our experiments with four real-world environments demonstrate that BPC-based enhancement approaches are promising pre-processing modules for speech recognition in practical settings. Our findings indicate that the perceptual loss approach, despite some compromises in subjective evaluation, yields better accurate recognition results than the Phone and baseline approaches. Meanwhile, the Phone/BPC-level E2E-ASR objective approaches strike a balance between subjective evaluation and recognition results, achieving improved performance across all evaluation metrics. Specifically, the BPC(M)-level E2E-ASR objective approach outperformed all the other approaches in all experiments.}
% From the four types of real-world environments, we can see the BPC-based enhancement approaches have the potential as pre-processing modules for speech recognition in practice. Our findings suggest that while the perceptual objective approach may have certain compromises in terms of subjective evaluation, it still yields the most accurate recognition results in non-stationary noise. On the other hand, the Phone/BPC-level E2E-ASR objective approaches enhance both subjective evaluation and recognition results, achieving a balance across all evaluation metrics. Finally, within the E2E-ASR objective approaches, the BPC(M)-level approach outperforms the Phone-level approach in all experiments.}

\begin{table}
\centering
\caption{WERs ({\%}) of three SE-E2E-ASR systems (Phoneme, BPC(M), and BPC(M)
Perceptual) and Baseline in different SNRs with four noise types on the TMHINT dataset.}
\label{tab:wer}
\begin{tabular}{ccccc}
\toprule
{SNR} & {Baseline } & {\makecell{BPC(M)-PL }}  & {Phoneme } & {BPC(M) }  \\
\midrule
-5 & 85.3\% & 81.4\% & 84.3\% & 82.6\% \\
-1 & 58.7\% & 53.7\% & 57.9\% & 55.4\% \\
1 & 44.3\% & 40.2\% & 42.3\% & 42.2\% \\
5 & 20.8\% & 19.2\% & 20.5\% & 19.5\% \\
\midrule
Avg. & 52.3\% & 48.6\% & 51.3\% & 49.9\% \\
% -7 &   92.8\% & 90.4\% & 91.9\% & 90.7\% 	\\
% -5 &   85.3\% &	81.4\% & 84.3\% & 82.6\%   	\\
% -3	& 73.1\%  &	69.6\% & 72.1\% & 70.5\%    \\
% -1	& 58.7\%  &	53.7\% & 57.9\% & 55.4\%   	\\
% \midrule
% Avg. & 77.5\% &	73.8\% & 76.6\% & 74.8\%  \\
\bottomrule
\end{tabular}
\end{table}

% \begin{table}
% \centering
% \caption{Relative WER with respect to the unprocessed speech in different SNRs and noise types on TMHINT experiments.}
% \label{tab:wer}
% \begin{subtable}{}
% \centering
% \caption{engine and street noises.}
% \label{tab:wer_es}
% \begin{tabular}{ccccc}
% \toprule
% {SNR} & {Baseline (\%)} & {Phone (\%)} & {BPC(M) (\%)} & {BPC(M)-Deep (\%)} \\
% \midrule
% -10 & -2.68 & -2.43 & -2.66 & -3.10 \\
% -7 & -0.19 & -1.66 & -3.96 & -5.48 \\
% -6 & 4.09 & 3.20 & 0.71 & -1.91 \\
% -5 & 4.27 & 2.19 & -0.85 & -4.50 \\
% \midrule
% Avg. & 1.37	& 0.33	& -1.69 & -3.75 \\
% \bottomrule
% \end{tabular}
% \end{subtable}
% \begin{subtable}{}
% \centering
% \caption{white and pink noises.}
% \label{tab:wer_wp}
% \begin{tabular}{ccccc}
% \toprule
% {SNR} & {Baseline (\%)} & {Phone (\%)} & {BPC(M) (\%)} & {BPC(M)-Deep (\%)} \\
% \midrule
% -10 & 7.97 & 7.66 & 7.02 & 7.34 \\
% -7 & 33.13 & 32.08 & 27.12 & 30.65 \\
% -6 & 49.83 & 49.86 & 39.59 & 44.74 \\
% -5 & 65.90 & 65.27 & 51.90 & 56.49 \\
% \midrule
% Avg. & 39.21	& 38.72	& 31.41	& 34.80 \\
% \bottomrule
% \end{tabular}
% \end{subtable}
% \end{table}

\subsubsection{Results for Speech Denoising and Dereverberation}
In addition to the denoising task, we evaluated the proposed methods on utterances further corrupted by reverberation, which is a more challenging task. For the training set, the clean utterances from the TMHINT corpus were first mixed with 104 noise types at 31 SNR levels from -10 to 20 dB and then distorted by reverberation. The test set had reverberated and noise-corrupted utterances, where seven unseen noise types were added at 14 SNRs (from -10 to 10 dB). We used the room impulse response (RIR) generator to create the reverberation, and room impulse responses were generated using the image method and applied to the noisy utterances. The reverberation time (T60) was randomly selected from 0.3, 0.6, and 0.9 s to generate impulse responses for the training data, and T60 was set to 0.4 s for the testing data. The RIR had a total of 4,096 samples. The receiver position was [2m, 1.5m, 2m], the source position was [2m, 3.5m, 2m], and the room dimensions were [5m, 4m, 6m]. The experiment aimed to remove both the noise and reverberation; thus, the clean data was set as the target when training the SE model.

\begin{table}[t]
\centering
\caption{Average PESQ and STOI scores for SE-E2E-ASR system with BPC(M) on TMHINT corpus with noise and reverberation.}
\label{tab:Reverb_Result}
% \begin{tabular}{|c|c|c|}
% \hline
\begin{tabular}{|c|c|c|c|}
\hline
\multicolumn{2}{|c|}{} & PESQ & STOI \\ \hline
\multicolumn{2}{|c|}{Noisy Reverberation} & 1.373 & 0.482 \\ \hline
\multicolumn{2}{|c|}{SE Baseline} & 1.284 & 0.536 \\ \hline
\multirow{2}{*}{SE-E2E-ASR} & Phoneme & 1.288 & 0.540 \\ \cline{2-4}
& BPC(M) & \textbf{1.384} & \textbf{0.555} \\ \hline
\end{tabular}

%  & PESQ & STOI \\ \hline
% Noisy Reverberation & 1.373 & 0.482 \\ \hline
% SE Baseline & 1.284 & 0.536 \\ \hline
% SE-E2E-ASR Phone & 1.288 & 0.540 \\ \hline
% SE-E2E-ASR BPC(M) & \textbf{1.384} & \textbf{0.555} \\ \hline
% \end{tabular}
\end{table}

\begin{figure}[htbp]
\centering
\subfigure[Clean]{
\includegraphics[width=4cm]{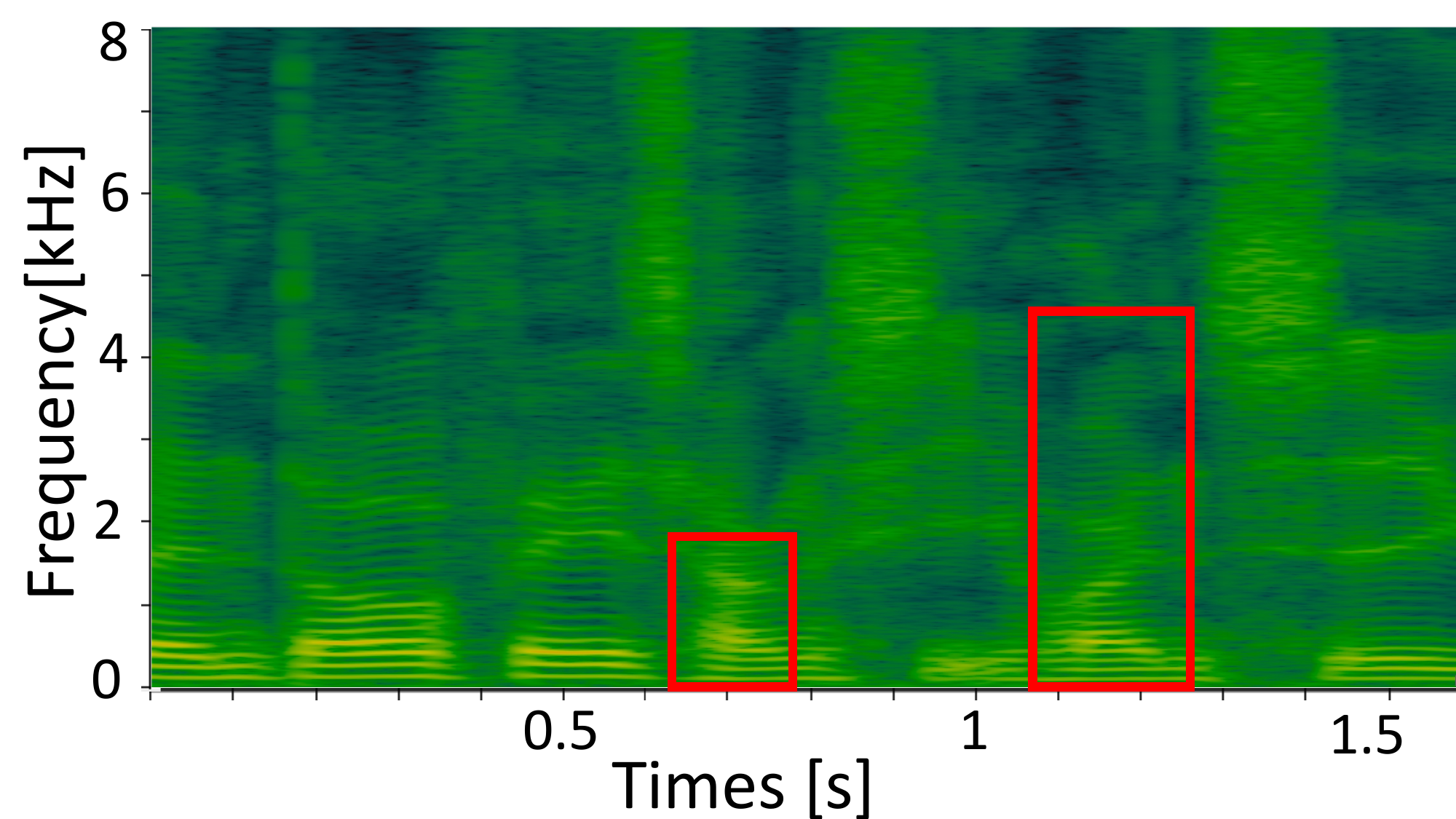}
}
% \quad
\subfigure[Noisy Reverberation]{
\includegraphics[width=4cm]{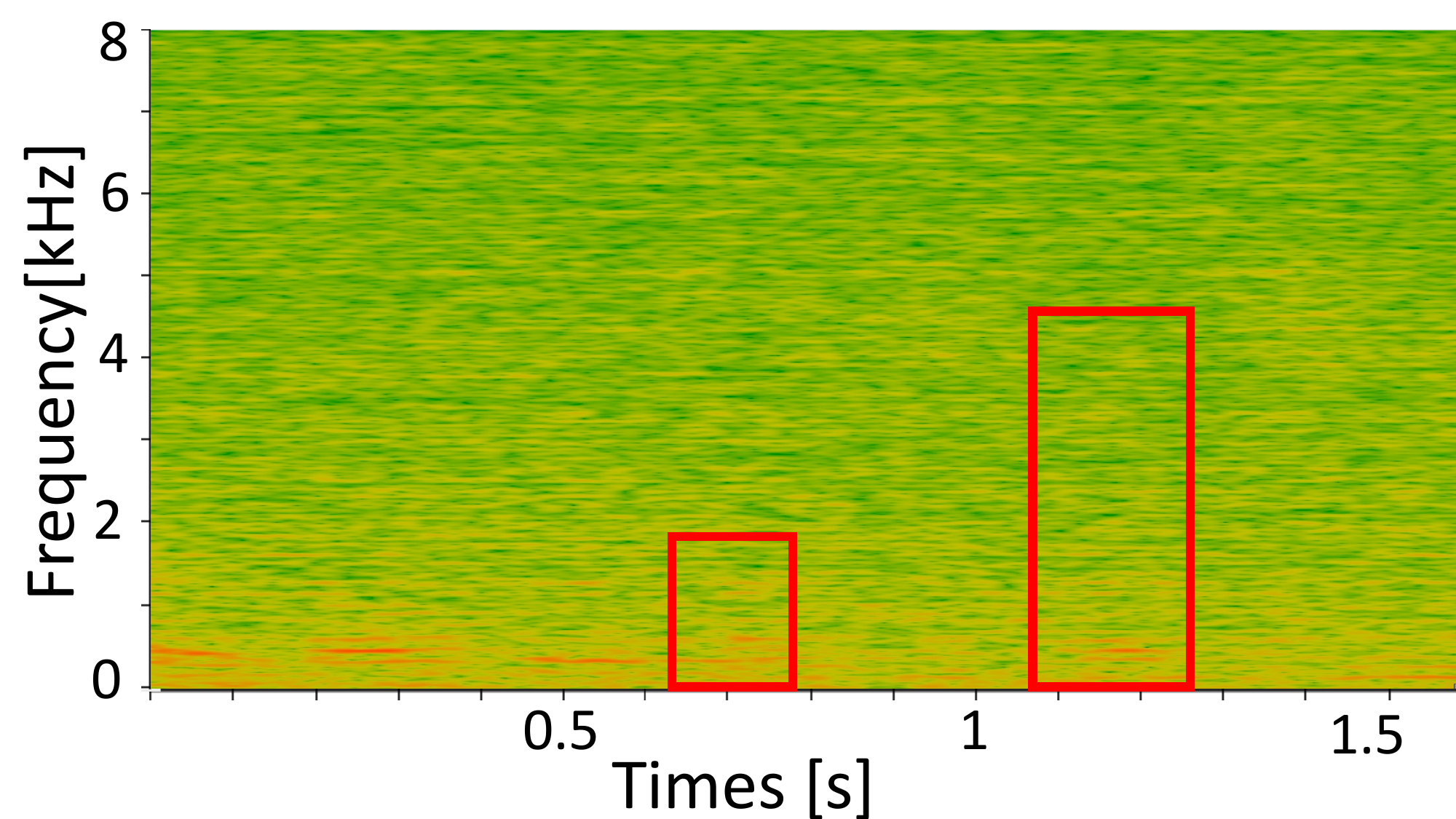}
}
\quad
\subfigure[SE Baseline]{
\includegraphics[width=4cm]{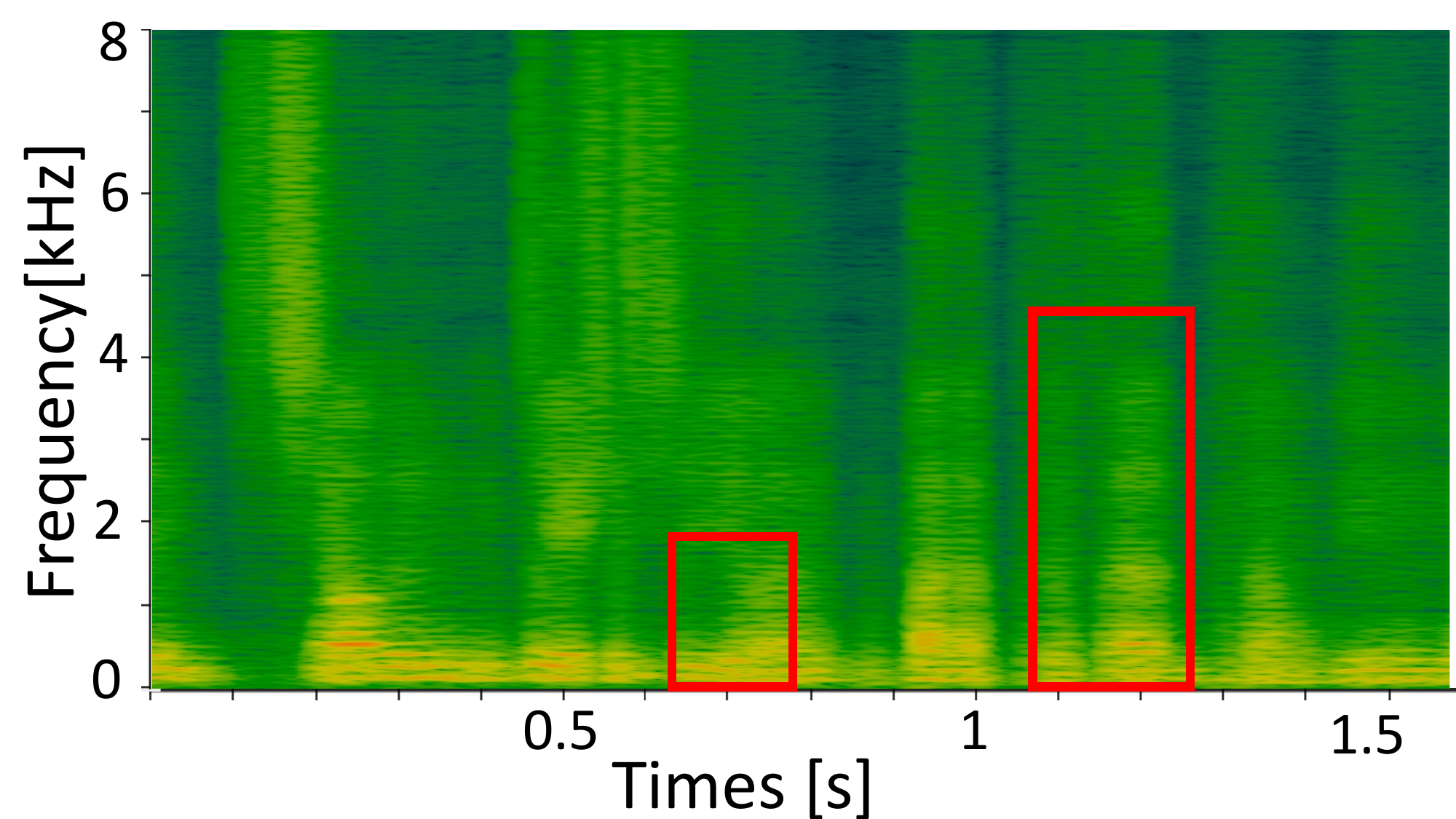}
}
% \quad
\subfigure[SE-E2E-ASR BPC(M)]{
\includegraphics[width=4cm]{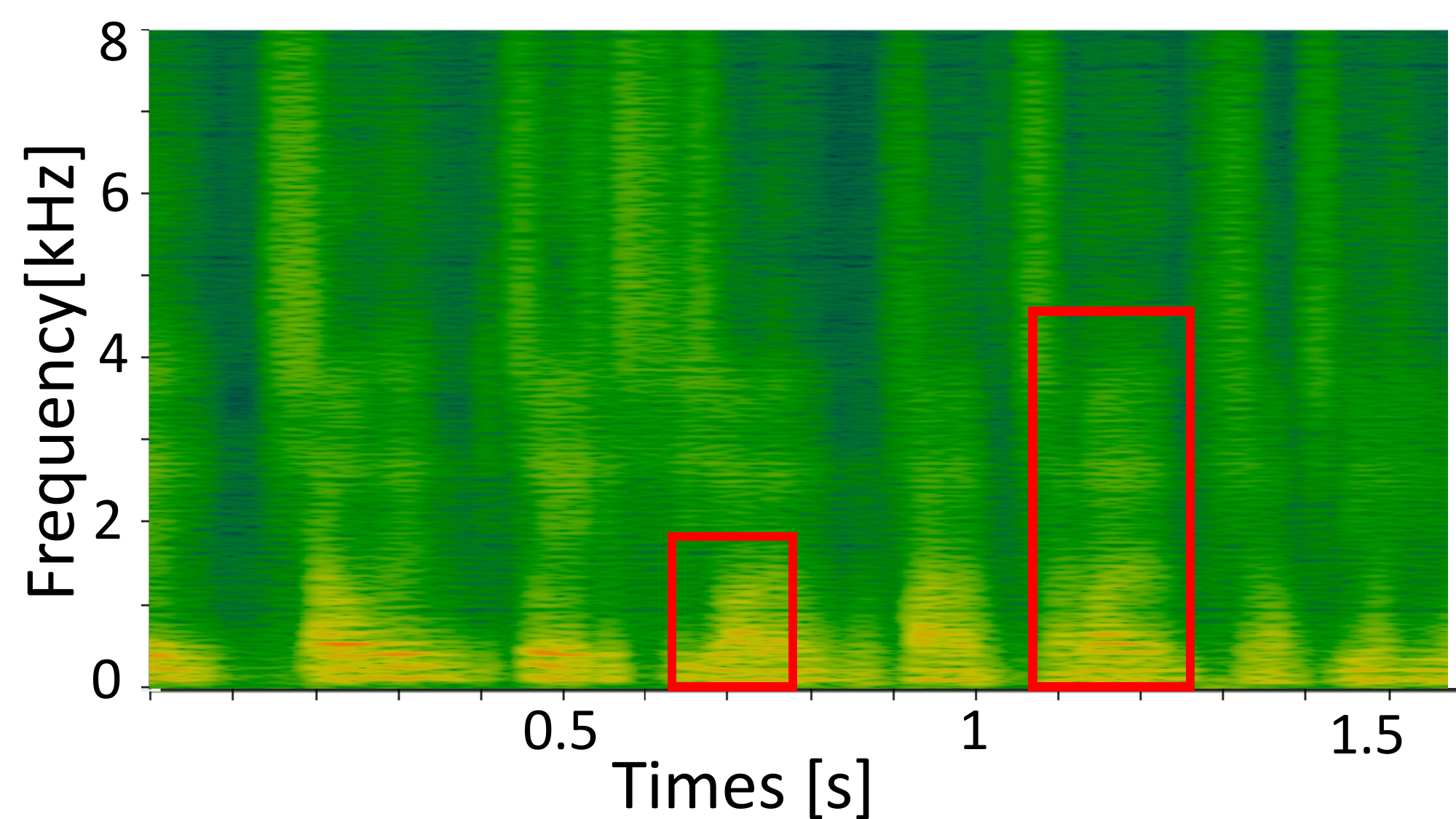}
}

\caption{Spectrogram plots of an utterance at different situations: (a) clean noise-rev-free, (b) noise-rev-corrupted, (c) noise-rev-corrupted and enhanced by the SE baseline , (d) noise-rev-corrupted and enhanced by the BPC(M)-level SE-E2E-ASR.}
\label{fig:tmhint_reverb_spectrogram}
\end{figure}

% Experiment discussion
Table \ref{tab:Reverb_Result} shows the average PESQ and STOI scores of the original reverberant noisy data and their two enhanced versions. It is observed from this table that the SE baseline improves STOI by 0.054 but worsens PESQ by 0.089. The phoneme-level objective provides slightly better but still degrades the PESQ score. According to \cite{han2015learning, zhao2018two}, a single DNN-based SE model may produce limited performance in a composite of noisy and reverberant condition: the STOI scores can be improved, but PESQ scores show no improvements, which matches our results. On the other hand, our proposed approach, BPC(M)-level SE-E2E-ASR objective, promotes both STOI and PESQ by 0.073 and 0.011, respectively, on average.  Therefore, we have demonstrated that the proposed system can effectively handle noise and reverberation issues simultaneously. 

The spectrograms of a clean utterance, its noisy-reverberant counterpart, and their enhanced signals are shown in Fig. \ref{fig:tmhint_reverb_spectrogram}. 
We highlighted two speech regions in the clean utterance and used them to compare the enhanced versions from the SE baseline and BPC(M)-level SE-E2E-ASR. We observe from this figure that the SE baseline does not recover the speech signals in the regions of the red box, whereas the speech is preserved for the proposed BPC(M)-level SE-E2E-ASR case.

% the enhanced speech shows improvement in both PESQ and STOI. The PESQ improves 0.1 in avarage, and the STOI improves 0.019, which is better than the denoising-only tasks.

\subsubsection{Results for Impaired SE}

The impaired utterances used in this study were based on the TMHINT corpus, which has 1,200 Mandarin Chinese utterances. We used non-impaired utterances from the TMSV corpus \cite{chuang2020improved} as the target speech data and dysarthric utterances as the input-impaired data to train the SE model. The TMSV corpus included 13 males and five female speakers, from which utterances from 13 male and four female speakers with better pronunciation accuracy were used in our task. The speech content in the TMSV corpus is similar to that in the TMHINT corpus. For each speaker, 240 utterances were used for training, 40 for validation, and 40 for testing. The 80-dimensional mel-spectrograms with 1024 DFT points and a 144-point frameshift for the utterances were extracted using the open-source ESPnet toolkit. The TTS-transformer model was pre-trained and used for the ISE task. Referring to Eq. (\ref{eqASR}), the transformer-based SE model was trained for 1000 epochs in advance with the SE loss ($\textbf{loss}_{SE}$) and then further trained for another 1000 epochs by adding the BPC(M)-level E2E-ASR loss ($\textbf{loss}_{ASR}$) to the SE loss. Parameter $\alpha$ was set to 0.001. Early stopping was implemented during training.

Subjective evaluations were conducted to evaluate the performance of impaired speech experiments. We randomly selected four utterances produced by 10 speakers (eight males and two females), which were processed by either the SE baseline or SE-E2E-ASR. A total of 21 subjects performed the evaluation, each given 120 samples (four samples $\times$ 10 speakers $\times$ three approaches) and requested to choose his/her preference among the three types of utterances. The listening test was conducted in a quiet environment, with an SNR level of approximately 55 dB. A Sennheiser HD599 headset was used, and the audio was played through a Samsung Tab S6 device. The resulting preference rates are presented in Table \ref{tab:sub_Dys}. It is surprising that the unprocessed utterances were least preferable at $16.8\%$, while both the SE baseline and SE-E2E-ASR with BPC(M) achieved more than twice the preference rate than the unprocessed speech, revealing the effectiveness of the SE models in amending impaired speech. In particular, the proposed SE-E2E-ASR obtained a $9.6\%$ higher preference rate than the SE baseline, showing that the multi-objective training of SE and BPC(M)-level E2E-ASR can further improve the speech quality of utterances converted from impaired speech.

% We conduct the subjective evaluation for the dysarthric experiments, we randomly choose four utterances in first 10 speaker targets (8 male speakers and 2 female speakers), comparing the preference among the unprocessed speech, VTN baseline, and E2E VTN-ASR BPC(M) results. There are totally 21 subjects to do the subjective evaluation, with 120 ($=$ 4 samples $\times$ 10 speaker targets $\times$ 3 approaches) evaluation samples per subject, totally 2520 preferences results. The preference for the three different methods are in the Table \ref{tab:sub_Dys}. We find out that the unprocessed speech are the least preferable, with only $16.8\%$, results from both VTN and E2E VTN-ASR-BPC(M) are higher than $30\%$, showing the VC models are effective in improving the speech quality. In addition, the preference of the E2E VTN-ASR-BPC(M) are $9.6\%$ higher than the VTN baseline, proving the BPC multi-objective training} method is able to further improve the speech quality for the converted speech signal.

\begin{table}[t]
\centering
\caption{Subjective preference evaluation for the impaired SE.}
\label{tab:sub_Dys}
\begin{tabular}{|c|c|}
\hline
 & Preference \\ \hline
Unprocessed & 16.8\% \\ \hline
SE Baseline & 36.8\% \\ \hline
SE-E2E-ASR BPC(M) & \textbf{46.4\%} \\ \hline
\end{tabular}
\end{table}

\section{CONCLUSION}
This study proposed a novel architecture that applies a BPC-based E2E-ASR to guide the SE process with contextual broad phonetic information to achieve superior speech quality and intelligibility. Three BPC clustering methods were investigated for the English corpus, and the evaluation results confirmed the context information of the BPC SE considerably over a wide range of SNR conditions. Furthermore, with the word-to-IPA transformation, we have extended the use of this novel approach to the Mandarin corpus with similar BPC clusters as in the English corpus experiments. Experimental results on three tasks, namely speech denoising, speech dereverberation, and impaired speech enhancement, verified the effectiveness of incorporating contextual broad phonetic information into SE to improve enhancement results. The main contributions of this study are as follows: (1) This is the first study that employed the context information of broad phonetic/articulatory phonemes classes for an end-to-end SE–ASR system. (2) We demonstrated that using both knowledge-based and data-driven BPCs as enhancement targets can further improve the quality and intelligibility of enhanced speech for both English and Mandarin. (3) We validated that knowledge-based BPCs are generally more versatile than data-driven BPCs and mono-phonemes, as they can be used in a wider range of scenarios. The main focus of this study is to examine the losses prepared by various pre-trained models, including AM and E2E-ASR systems, to leverage the SE performance. Our experimental results have validated the efficacy of including contextual broad phonetic information in SE training. In the future, we will further utilize the findings of this study to enhance other DL-based SE models.

% We validated that knowledge-based BPCs benefit SE more \textcolor{blue} {general} than data-driven BPCs and mono-phonemes. 
%\label{sec:typestyle}

% References should be produced using the bibtex program from suitable
% BiBTeX files (here: strings, refs, manuals). The IEEEbib.bst bibliography
% style file from IEEE produces unsorted bibliography list.
% -------------------------------------------------------------------------
%\setlength{\bibspacing}{0\baselineskip}
\footnotesize
\bibliographystyle{IEEEbib}
%\ninept
\bibliography{strings,refs}
\end{document}